\documentclass[aps,prl,reprint,superscriptaddress,floatfix]{revtex4-2}
\usepackage{amssymb}
\usepackage{physics}
\usepackage{graphicx}
\usepackage{mathtools}
\usepackage{amsmath}
\usepackage{amsthm}
\usepackage{amsfonts}
\usepackage[T1]{fontenc}
\usepackage{array}
\usepackage{multirow}
\usepackage{xcolor}
\usepackage{esint}
\usepackage{bm}
\usepackage{color}
\usepackage{bbm}
\usepackage{hyperref}
\usepackage{babel}
\usepackage{titlesec}
\usepackage{bbm}
\usepackage[inline]{trackchanges}
\newcommand{\lettersection}[1]{\textbf{#1}.---}

\usepackage{hyperref}
\hypersetup
{	colorlinks,%
	citecolor=green,%
	linkcolor=red,%
	urlcolor=blue%
}
\allowdisplaybreaks[4]

\newcommand{\be}{\begin{equation}}
\newcommand{\ee}{\end{equation}}
\newcommand{\ba}{\begin{aligned}}
\newcommand{\ea}{\end{aligned}}

\newcommand{\rket}[1]{\left|{#1}\right)}
\newcommand{\rbra}[1]{\left({#1}\right|}
\newcommand{\rbraket}[1]{({#1})}

\newcommand*{\TextVCenter}[1]{%
  \text{$\vcenter{\hbox{#1}}$}%
}

\newcommand{\LLG}{LLG}

\newcommand{\vlc}{v_{\mathrm{LC}} }

\newcommand{\BigotimesB}[2]{\bigotimes^{#1}_{\text{#2}}}
\newcommand{\Bigodot}[2]{{\bigodot^{#1}}_{\text{#2}}}
\newcommand{\BigodotB}[2]{\bigodot^{#1}_{\text{#2}}}
\newcommand{\Otimes}[2]{{\otimes^{#1}}_{\text{#2}}}

\newcommand{\UU}{\mathcal{U}}

\begin{document}
\title{Out-of-time-order correlator, many-body quantum chaos, \\
light-like generators, and singular values}

\author{Ke Huang}
\affiliation{Department of Physics, City University of Hong Kong, Kowloon, Hong Kong SAR, China}

\author{Xiao Li}
\email{xiao.li@cityu.edu.hk}
\affiliation{Department of Physics, City University of Hong Kong, Kowloon, Hong Kong SAR, China}

\author{David A. Huse}
\affiliation{Department of Physics, Princeton University, Princeton, New Jersey 08544, USA}

\author{Amos Chan}
\email{amos.chan@lancaster.ac.uk}
\affiliation{Physics Department, Lancaster University, Lancaster, LA1 4YW, United Kingdom}

\date{\today}
\begin{abstract}
We study out-of-time-order correlators (OTOCs) of local operators in 
spatial-temporal invariant or random 
quantum circuits using light-like generators (LLG) --- many-body operators that exist in and act along the light-like directions.
We demonstrate that the OTOC can be approximated by the leading singular value of the LLG, which, for the case of generic many-body chaotic circuits, is increasingly accurate as the size of the LLG, $w$, increases.
We analytically show that the OTOC has a decay with a universal form in the light-like direction near the causal light cone, as dictated by the sub-leading eigenvalues of LLG, $z_2$, and their degeneracies.
Further, we analytically derive and numerically verify that the sub-leading eigenvalues of LLG of  any size can be accessibly extracted from those of LLG of the smallest size, i.e., $z_2(w)= z_2(w=1)$.  
Using symmetries and recursive structures of {\LLG}, we propose two conjectures on the universal aspects of generic many-body quantum chaotic circuits, one on the algebraic degeneracy of eigenvalues of LLG, and another on the geometric degeneracy of the sub-leading eigenvalues of LLG.
As corollaries of the conjectures,  we analytically derive the asymptotic form of the leading singular state, which in turn allows us to postulate and efficiently compute a product-state variational ansatz away from the asymptotic limit. 
We numerically test the claims with four generic circuit models of many-body quantum chaos, and contrast these statements against the cases of a dual unitary system and an integrable system.
\end{abstract}
\maketitle

\begin{figure*}[t]
\centering
\includegraphics[width=2\columnwidth]{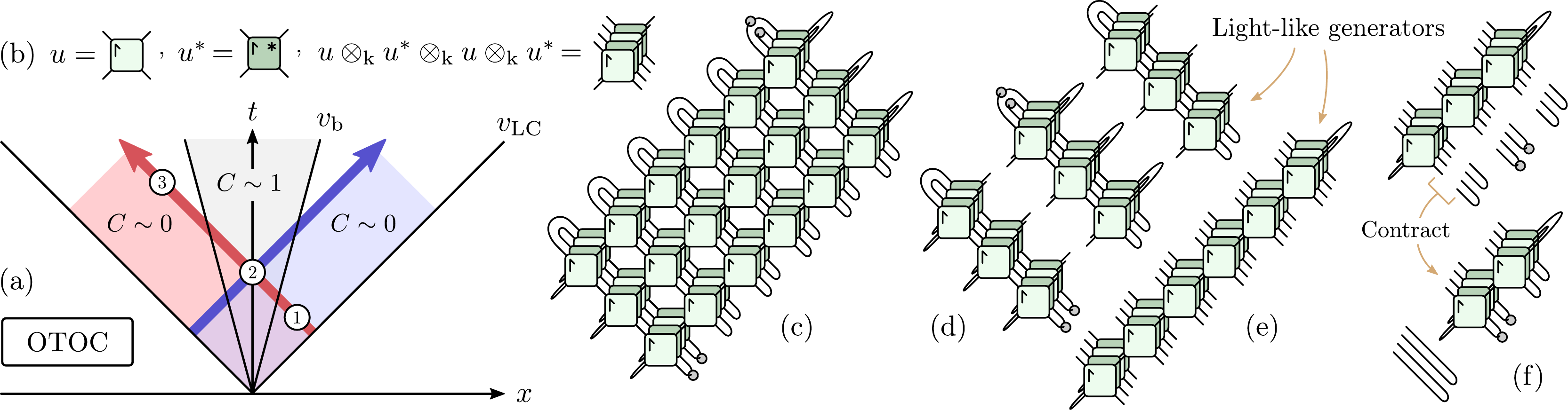}
\caption{\label{Fig:otoc}
(a) Illustration of the behaviour OTOC $\tilde{C}(x,t) = C(w,\tau)$ in the $x$ and $t$ plane (or $w$ and $\tau$ plane). 
${C}\sim 1$ sufficiently within region 2, the butterfly cone, defined by the butterfly velocity $v_{\text{b}}$~\cite{keyserlingk2018, nahum2018}, in which the operator spreads. 
${C}\sim 0$ in regions 1 and 3 outside the above region but inside the system's causal light cone defined by the light cone velocity $\vlc$. 
The {\LLG} can be used to generate OTOC along the red arrow, which goes through the regime ${C}\sim 0$, ${C}\sim 1$ and ${C}\sim 0$ in sequence, the ``010-behavior'', implying non-trivial properties of the generators. 
(b) The diagrammatical representation of unitary gates and their conjugates. 
(c) The representation of OTOC $ \tilde{C}(x=2,t=8) =C(w= 6,\tau=3) $ after the annihilation of unitaries and their complex conjugates outside the light cone. The grey circles represent the (generalization of) Pauli operators.
The OTOC can be generated by (d)
$(\tilde L_\tau |$,  
$ | \tilde R_\tau)$, and the right-moving {\LLG}; or the (e) left-moving many-body {\LLG}.
(f) {\LLG} of size $w$ can be reduced to {\LLG} of size $w' < w$ upon contraction with certain states.
}
\end{figure*}

\lettersection{Introduction}
A generic interacting {quantum} many-body  system obeys the eigenstate thermalization hypothesis (ETH) and evolves towards the local equilibrium state~\cite{Deutsch, Srednicki, Rigol2008}. 
During this process, information of local perturbation spreads among non-local degrees of freedom of the system, and can be diagnosed using the out-of-time-order correlator (OTOC) \cite{MSS, shenker2014black, LarkinOvchinnikov}, a correlation function between multiple operators probed at times which are out of order~\cite{li2017measuring, garttner2017measuring, landsman2019verified, joshi2020quantum, blok2021quantum, mi2021information, braumuller2022probing, Hashimoto_2017, galitski2017, luitz2017, motrunich2018otoc, Garc_a_Mata_2018, richter2018, Pappalardi_2018, swingle2018unscrambling, dag2019, Ch_vez_Carlos_2019, xu2020, znidaric2022, Gu_2022, Zhou_2023, chenu2023unveiling, santosrigol2023}. 
Recent developments of random quantum circuits as toy models of spatially-extended many-body quantum chaotic systems have led to discoveries of a plethora of novel phenomena~\cite{Nahum2017, keyserlingk2018, nahum2018, rakovszky2018, khemani2018, cdc1, cdc2,  bertini2018exact, bertini_exact_2019, friedman2019, nahum2019statmech, bertini_op_2020, Li_2018, Skinner_2019, Li_2019, Gullans_2020, altman2019, Jian_2020, Zabalo_2020}. 
In particular, the Floquet operator of a given generic quantum many-body circuit, a \textit{space-like generator}, captures novel signatures in spectral correlation~\cite{cdc2, saad2019semiclassical, Gharibyan_2018, roy2020sff, moudgalya2021} and eigenstate correlation~\cite{cdc3, Foini2019ethotoc} beyond those described in the random matrix theory and ETH; while its associated space-time dual transfer matrix, a \textit{time-like generator},  have been found {to} exhibit non-unitary dynamics~\cite{guhr2016kim, cdc2, bertini2018exact, waltner2019dual, friedman2019, cdclyap, garratt2020dw, lerose2020influence, ippoliti2021fractal, grover2021entanglement}, and remarkably belongs to the universality class of the Ginibre ensemble~\cite{Chan2022, Shivam2023}. 
\textit{Light-like generators} ({\LLG}s)
\cite{bertini_op_2020, kos2020correlations, claeys2020duotoc_vmax, bertini2020otoc, Claeys2022integrable, claeys2023duotoc}
 are many-body operators that exist in and act along the LL directions [Fig.~\ref{Fig:otoc}(d-e)].
There are at least two reasons why the study of LLGs is interesting. Firstly, correlation functions of local operators, e.g., two-point correlation function and OTOC, in quantum circuits can naturally be written in terms of LLGs due to the upper bound on the propagation speed of quantum information as given by the causal light cone.
Secondly, as we will argue and demonstrate below, LLG captures universal signatures of many-body quantum dynamics. 
As a heuristic, along the LL direction, OTOC can be generated from a \LLG\ of a fixed length, and takes the values around 0, 1, and 0 sequentially, exhibiting the ``010''-behavior for generic chaotic circuits [Fig.~\ref{Fig:otoc}(a)]. 
Such phenomena suggest that the full behavior of OTOC cannot be dictated by the leading eigenvalues of {\LLG}s, and in fact, the non-normal nature of {\LLG}s underlies the universal features of OTOC.

\lettersection{Models}
We consider many-body quantum systems modelled by unitary circuits of two-site gates arranged in the brick-wall geometry, $U(t)=\prod_{s=1}^t \bigotimes_{\substack{i \in 2 \mathbb{Z} + s\, \text{mod}\,2}}u_{i,i+1}$, where $u_{i,i+1}$ are $q^2$-by-$q^2$ unitary operator acting on site $i$ and $i+1$, see Fig.~\ref{Fig:otoc}.    
{These two-site gates} are selected to model three different dynamics without local conserved quantities, {including:} (i) four models displaying generic many-body quantum chaos, namely the Haar-random (HRM), random phase (RPM), three-parameter (3PM) models, and a variant of HRM with $Z_2$ and time-reversal symmetry ($Z_2$-COE) [to compare with (iii) below]; (ii) a DU model,
and (iii) an integrable model called the XYZ circuit (XYZc). 
The precise definitions of these models are provided in supplementary material (SM) \cite{supplementary} and are not important to the derivations of results below.  
We study mainly two symmetry classes, namely the spatial-temporal invariant circuits whose gates are all identical, and the spatial-temporal random circuits whose gates are independently chosen from certain ensemble. Ensemble averages are taken for the spatial-temporal random circuits but not for the invariant ones.

\lettersection{Light-like generators}
The OTOC is defined as
\begin{align}
	\tilde C(x,t)&:=\frac12\expval{[\sigma(0,t),\sigma(x,0)]^\dag[\sigma(0,t),\sigma(x,0)]}\nonumber\\
	&=1-\expval{\sigma(0,t)\sigma(x,0)\sigma(0,t)\sigma(x,0)},
\end{align}
where $\sigma(x,t)= U(t) \sigma_x [U(t)]^\dagger$, and $\sigma_x$ is a hermitian and unitary local operator acting on site $x$. 
To represent OTOC, it is convenient to define a tensor product of unitary gates in the replicated space with $\mathcal{U}_2 \equiv u \;  \Otimes{}{k}  \; u^* \; \Otimes{}{k} \;  u \; \Otimes{}{k} \;  u^*$ [Fig.~\ref{Fig:otoc}(b)] and states in the replicated space with 
\begin{equation}
\begin{aligned}
& \left| 0_{\sigma,\mu} \right\rangle := \frac1{\sqrt q}
\hspace{-0.06cm}
 \TextVCenter{ \includegraphics[width=0.6cm]{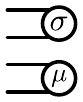} }
\hspace{-0.09cm}
\,
,  \; \; 
 &\left| 1_{\sigma,\mu} \right\rangle := \frac1{\sqrt q} 
\hspace{-0.06cm}
\TextVCenter{ \includegraphics[width=0.8cm]{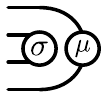} }
\hspace{-0.09cm}
 \, ,
\end{aligned}
\end{equation}
where $\sigma, \mu$ are identities or the (generalization of) Pauli matrices. 
We further define $\ket{0} \equiv \ket{0_{\mathbbm{1},\mathbbm{1}}} $ and $\ket{1} \equiv \ket{1_{\mathbbm{1},\mathbbm{1}}}$.
By annihilating pairs of unitary gates and their conjugates outside of the light cone, as illustrated in Fig.~\ref{Fig:otoc}(c), %
the OTOC can also be expressed in terms of {\LLG}s, 
\begin{align}\label{Eq:OTOC_LL}
	C(&w,\tau) := \tilde C(x= w-\tau-1,t = w+\tau-1)\nonumber\\
	&= 1-\mel{L_w}{T_{\text{L},w}^\tau}{R_w}
	= 1-\rbraket{\tilde L_\tau | T_{\text{R},\tau}^{w-2} | \tilde R_\tau},
\end{align}
where $w=(t + x)/2 + 1$ and $\tau = (t - x)/2$ are the shifted light cone coordinate labelling the diagonal width and length of the overlapped light cone region of the two local operators in the OTOC [see Fig.~\ref{Fig:otoc}(c)].
The left-moving and right-moving {\LLG}s are, $T_{\text{L},w}=\rbra{0}\BigodotB{w}{r}\UU_2\rket{1}$ and $T_{\text{R},w}=\bra{1}\BigodotB{w}{l}\UU_2\ket{0}$, as illustrated in Fig.~\ref{Fig:otoc}(e) and~\ref{Fig:otoc}(d).
$\Bigodot{}{r}$ ($\Bigodot{}{l}$) denotes the contraction between the northeast (northwest) legs of a tensor with the southwest (southeast) leg of another tensor. Similarly, angle (round) brakets denote vectors with open legs along the left-going (right-going) LL direction. 
The initial and final states associated with the left-going {\LLG} are given by
$\bra{L_w}=\bra{1^{w-1}}\Otimes{}{}\bra{1_{\sigma,\sigma}}$, and $\ket{R_w}=\ket{0_{\sigma,\sigma}}\Otimes{}{}\ket{0^{w-1}}$, where we have defined $\bra{1^w}:=\BigotimesB{w}{r}\bra{1}$ and similarly for the other states.
The initial and final states associated with the right-going {\LLG} are 
$({\tilde L_w}|=\rbra{0^{w}}\qty[\bra{1}\BigodotB{w}{l}\UU_2\ket{0_{\sigma,\sigma}}],$ and 
$|{\tilde R_w})=\qty[\bra{1_{\sigma,\sigma} }\BigodotB{w}{l}\UU_2\ket{0}]\rket{1^w}$.
For succinctness, the subscript $\text{R, L}$ of the {\LLG} will be hidden if the discussion applies to both directions.
One can prove that the moduli of all  eigenvalues of $T_w$ are equal or less than 1 \cite{supplementary}. An obvious pair of left and right eigenstates of $T_{w}$ with eigenvalue $1$ is $\bra{1^w},\ket{0^w}$, normalized such that $\braket{1^w}{0^w}=1$.
Generally, we find that these are the only eigenstates with eigenvalue $1$ except for the DU case, i.e., the subleading eigenvalue of $T_w$ satisfies $\abs{z_2(w)}<1$. 
Hence, if we remove the leading eigenvalue from $T_w$ and define $F_w:=T_w-\dyad{0^w}{1^w}$, we can write 
\begin{align}
	C(w,\tau)=-\mel{L_w}{F_{\text{L},w}^\tau}{R_w}=-\rbraket{\tilde L_\tau | F_{\text{R},\tau}^{w-2} | \tilde R_\tau}. 
\end{align}
Here, we have used $F_w^\tau=T_w^\tau-\dyad{0^w}{1^w}$ and $\braket{L_w}{0^w}=\braket{1^w}{R_w}=1$. 

$F_w$ and $T_w$ are both nonnormal matrices, so their operator norms are not solely determined by their eigenvalues. 
It turns out that their operator norms satisfy~\cite{supplementary}
\begin{align}
	\lim_{w\to \infty}\norm{T_w}=q^\alpha,\quad \lim_{w\to\infty}\frac{\norm{F_w}}{q^w}=1,
\end{align}
where $0 \leq \alpha \leq 1$ is independent of $w$. 
The second relation is a direct result of the $w$-independence of $\alpha$, because $F_w=-\dyad{0^w}{1^w}+T_w$ and the first term dominates the operator norm.
If the subleading eigenvalue $z_2(w)$ of $T_w$ is less than $1$, the large-$\tau$ behavior of the {\LLG}\ is described by~\cite{supplementary}
\begin{align}\label{Eq:Largetau}
\lim_{\tau\to\infty}\norm{F_w^\tau}=0, \quad 
\lim_{\tau\to\infty}\norm{T_w^\tau}=q^w. 
\end{align}
Moreover, as $\norm{F_w^\tau}$ is still dominated by $\dyad{0^w}{1^w}$ for $\tau<w/\alpha$, there is a plateau at $\norm{F_w^\tau}\approx q^w$ before the inevitable decay, which differentiates $F_w$ from the normal matrices.
The asymptotic behaviors of {\LLG} for the cases of generic chaos and DU (discussed below) are summarized in Table.~\ref{Tab:LL}. 
\begin{table}[t]
\caption{\label{Tab:LL} Summary of properties of the {\LLG}s of generic quantum many-body chaotic circuits, and DU circuits.}
\begin{ruledtabular}
\begin{tabular}{ccc}
  & Generic Chaos & Dual Unitary\\
\hline
Block Structure & Upper-triangular & Diagonal \\
\hline
Lead. sing. val.  & $q^w~ (\tau\gg w)$ & $1$ \\
of $T_w^\tau$ & $q^\alpha ~(\tau=1,w\gg1)$ & \\
 \hline
Lead. eig. val.  & $z_2(w)=z_2(w=1)<1$ & $z_2(w)=1$ \\
 of $F_w$ & $a(z_2)\stackrel{\text{conj}}{=}w$ & $a(z_2)\geq w$ \\
  & $g(z_2)\stackrel{\text{conj}}{=}1$ & $g(z_2)=a(z_2)$ \\
 \hline
Lead. sing. val. & $\tau^{w-1}z_2^\tau~ (\tau\gg w)$ & $q^w$\\
of $F_w^\tau$  & $q^w ~(\tau\lesssim w,w\gg1)$ &  \\
 \hline
Lead. sing. state& $\ket{1^{w-1}}\otimes\ket{z_2}~(\tau\gg w)$ & Exactly solvable   \\
  of $F_w^\tau$  & $\ket{1^w} ~(\tau\lesssim w,w\gg1)$ & \& $\tau$-independent \\
\end{tabular}
\end{ruledtabular}
\end{table}

\lettersection{Recursivity and symmetries}
For systems with translation symmetries in space and in time, the {\LLG} of a given size can be reduced to smaller ones, 
\begin{align}\label{Eq:Reduce}
	&F_w\ket{0^{m}}\otimes\ket{\psi}=\ket{0^{m}}\otimes\qty(F_{w-m}\ket{\psi}),\nonumber\\
	&\bra{\psi}\otimes \bra{1^{m}}F_w=\qty(\bra{\psi}F_{w-m})\otimes \bra{1^{m}}, 
\end{align}
where $\ket{\psi}$ is an arbitrary state. 
Hence, we can decompose the Hilbert space into four orthogonal subspaces, $P_{0\bar1},P_{\bar0\bar1},P_{01},P_{\bar01}$ where the subscript $0$ and $1$ denote the states with $\ket{0}$ on the leftmost site of the {\LLG} and $\ket{1}$ on the rightmost one, and the bar denotes that the state on the leftmost (rightmost) site is orthogonal to $\ket{0}$ ($\ket{1}$). 
Note that the irreducible subspace is $P_{\bar01}+P_{\bar0\bar 1}$ for the ket states, and $P_{0\bar 1}+P_{\bar0 \bar 1}$ for the bra states.
In terms of these four sectors, the {\LLG} {for $w\geq 2$} becomes block-upper-triangular, schematically illustrated by
\begin{align}\label{Eq:Decomposition}
	\hspace{-0.06cm}
 \TextVCenter{ \includegraphics[width=6cm]{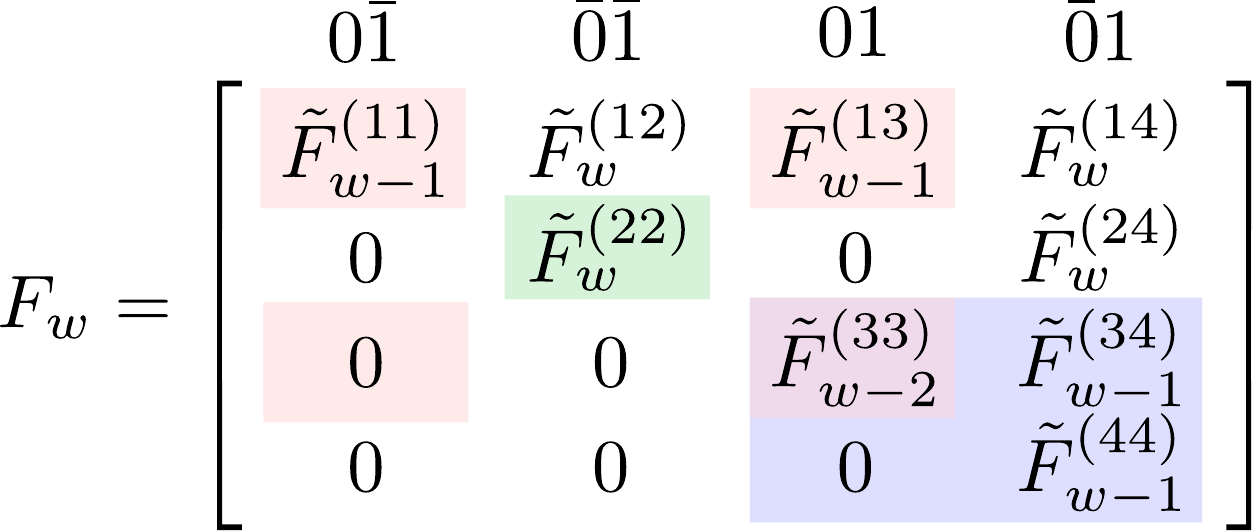} }
\hspace{-0.09cm} \, ,
\end{align}
where $\tilde{F}_{w}^{(ij)}$ is a tensor obtained from contracting a {\LLG} of size $w$, and the superscripts of $\tilde{F}_{w}^{(ij)}$ denote the position of the block in the matrix. 
In particular, we have $\tilde{F}_{w=0}^{(33)} = 0$. 
Hence, the eigenvalues of $F_w$ are determined by the diagonal blocks. 
We can further show that the blocks in red and blue separately constitute a {\LLG} of width $w-1$, that $\tilde{F}_{w-2}^{(33)}$ is identical to $F_{w-2}$, and that $\tilde{F}_w^{(22)}$ in green is a block that appears in $F_w$, but not in smaller {\LLG}~\cite{supplementary}. 
Consequently, we can obtain recursively for $w \geq 2$ that
\begin{align}
	a(z,w)=2a(z,w-1)-a(z,w-2)+\tilde{a}(z,w), \label{Eq:Recursion}
\end{align}
where $a(z,w)$ and $\tilde{a}(z,w)$ are respectively the algebraic multiplicity of each eigenvalue $z$ of $F_w$ and $\tilde{F}_w^{(22)}$. 
In particular, when $z$ is not an eigenvalue of $F_w$ ($\tilde{F}_w^{(22)}$), we set $a(z,w)= 0$ ($\tilde{a}(z,w)=0$). 
We further define $a(z,w=0)\equiv0$.
Among these eigenvalues, the leading eigenvalue $z_2(w)$ of $F_w$ is particularly important, as it partially determines the decay rate along the LL direction (see Eq.~\eqref{eq:otoc_decay} below). 
Under mild assumptions, one can prove that the asymptotic decay rate of $C(w,\tau)$ is limited by the decay rate of $F_{w=1}$, i.e., all eigenvalues $z$ of $F_w$ satisfy $\abs{z} \leq \abs{z_2 (w=1)}$, and hence there always exists ~\cite{supplementary}
\begin{align}\label{Eq:z2}
	z_2(w)=z_2(w=1),
\end{align}
regardless of whether the system is integrable or chaotic.  
In other words, $z_2(w)$, and hence the OTOC decay rate Eq.~\eqref{eq:otoc_decay}, are generally accessible through \LLG\ of width $w=1$. 
Furthermore, we observe that the \LLG\ generically has only one leading eigenvalue, and therefore $a(z_2,w=1)=1$. 
These statements are rigorously proved for the ensemble-averaged {\LLG} for the spatial-temporal random HRM~\cite{supplementary}.
If the system is chaotic, it is natural to assume that the eigenvalues of the new block $\tilde{F}_w^{(22)}$ are not fine-tuned, i.e., they are not identical to those of any smaller {\LLG}s. 
Hence, we have \textbf{Conjecture~1 on  circuit models of generic many-body quantum chaos}: if $z$ is an eigenvalue of $F_{w-1}$, then 
 \begin{equation} \label{Eq:ConjectureA}
\tilde{a}(z,w)=0 \;. 
\end{equation}
Particularly, we have $\tilde{a}[z_2(w),w]=0$.
 Furthermore, the geometric multiplicity $g[z_2(w),w]$ generally does not increase and remains to be $1$ because the off-diagonal blocks are not expected to be zero for generic chaotic models. 
Hence, we introduce \textbf{Conjecture~2 on  circuit models of generic many-body quantum chaos}: 
 \begin{equation} \label{Eq:ConjectureG}
 g[z_2(w),w]=1 \;.
 \end{equation} 
Consequently, the {\LLG}\ is not only non-normal, but generally non-diagonalizable. 
This reducibility also suggests that the OTOC is described by some smaller {\LLG}s $F_{w-m}$ if one of the initial states is reducible, while only the OTOC with irreducible initial states, such as $\bra{L_w},\ket{R_w}$, is truly governed by $F_w$. Note that $(\tilde L_w|,|\tilde R_w)$ contain both irreducible and reducible components which are controlled by {\LLG}s of different sizes.

\begin{figure}[t]
\includegraphics[width=0.95\columnwidth]{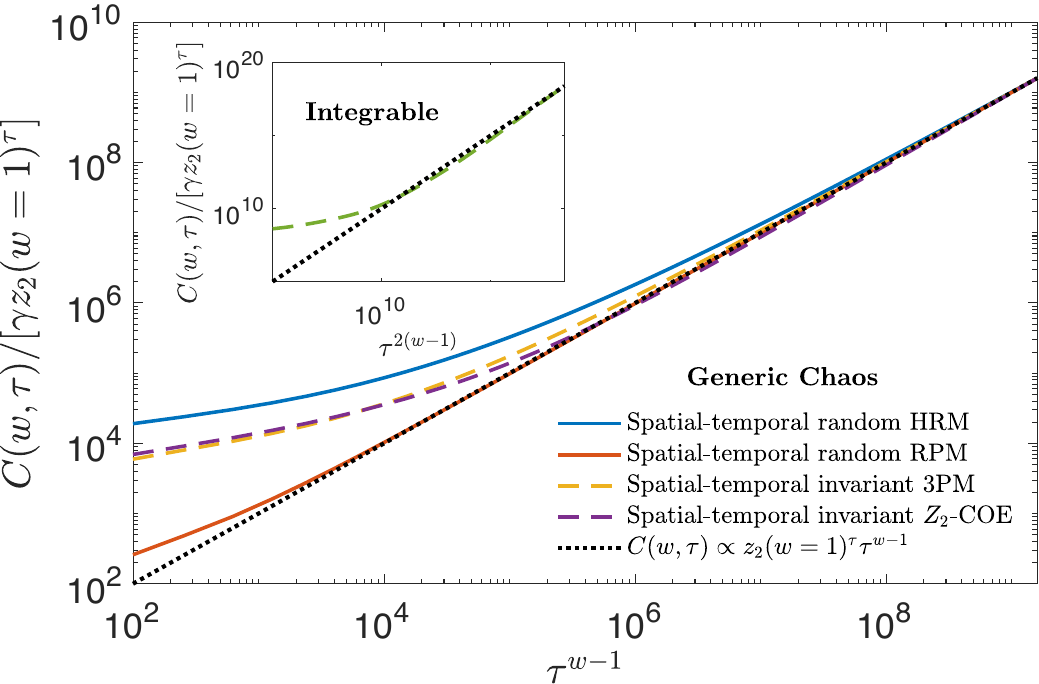}
\center
\caption{\label{Fig:OTOC_tail} 
Verifications of Eq.~\eqref{Eq:z2} and Conjectures 1 and 2. 
The large $\tau$ behavior of the OTOC for four generic chaotic circuit models. 
The inset shows an integrable model, the XYZc model, with a distinct scaling behavior. 
Here, we take $w=5$, and $\gamma$ is a non-universal model-dependent constant derived from fitting. }
\end{figure}

\lettersection{Large $\tau$ behavior}
In the large $\tau$ limit, the OTOC eventually decays according to \cite{supplementary},
\begin{align}\label{eq:otoc_decay}
C(w,\tau)\propto \lambda_{w,\tau}\propto\tau^{\varphi[z_2(w),w]}z_2(w)^\tau,
\end{align}
where $\varphi(z,w)=N_{\max}(z,w)-1$, and $N_{\max}(z,w)$ is the dimension of the largest Jordan block corresponding to eigenvalue $z$. 
From the two conjectures, %
and the empirical result $a(z_2,w=1)=1$, we have, for generic chaotic circuit models, 
\begin{equation}
\varphi(z,w)=a(z,w)-1 = w-1 \;.
\end{equation}
For the spatial-temporal random HRM, we can prove that $z_2(w)=q^2/(q^2+1)$ and $\varphi[z_2(w),w]=w-1$, as consistent with the conjectures~\cite{supplementary}.
To further test the conjectures, we numerically calculate four chaotic circuit models. 
For the spatial-temporal random case, we test the HRM and RPM after ensemble averages, and for the spatial-temporal invariant case, we test the 3PM and the $Z_2$-COE model. 
All four models agree with our conjecture excellently, as shown in Fig.~\ref{Fig:OTOC_tail}. 
The results also confirm that the {\LLG}s are generally non-diagonalizable, and thus, one cannot even numerically obtain all eigenvalues accurately. 
For the integrable model XYZc, the conjectures are indeed violated as shown by the inset of Fig.~\ref{Fig:OTOC_tail}. Numerically, we find that the leading eigenvalue of $F_{w}^{(22)}$ is $z_2(w=1)$, resulting in much larger Jordan blocks. 
For the XYZc, we empirically find that $\varphi[z_2(w), w]=2(w-1)$, as verified in Fig.~\ref{Fig:OTOC_tail}.

\lettersection{Leading singular value approximation}
Under Conjecture~2, there is a unique 
pair of leading singular states that dominates the \LLG s in the large $\tau$ limit~\cite{supplementary},
\begin{align}
	\lim_{\tau\to\infty}\norm{\lambda_{w,\tau}^{-1}F_{\text{L},w}^\tau-\dyad{\lambda_{w,\tau}^\text{L}}{\lambda_{w,\tau}^\text{R}}}=0,
\end{align}
where $\lambda_{w,\tau}$ is the leading singular value of $F_{\text{L},w}^\tau$ with its left and right singular states $\bra{\lambda_{w,\tau}^\text{L}},\ket{\lambda_{w,\tau}^\text{R}}$. 
Thus, the OTOC is exactly controlled by the leading singular value in this limit. 
Therefore, we introduce the leading singular value approximation (LSVA) for the left-moving {\LLG},
\begin{align}
	C_{\text{LSVA}}(w,\tau):=-\lambda_{w,\tau}\braket{L_w}{\lambda_{w,\tau}^\text{L}}\braket{\lambda_{w,\tau}^\text{R}}{R_w},
\end{align}
and conjecture that the LSVA describes the generic behavior of the OTOC. 
The $C_\text{LSVA}$ of the right-moving {\LLG} can be similarly defined. 

\begin{figure}[t]
\centering
\includegraphics[width=\columnwidth]{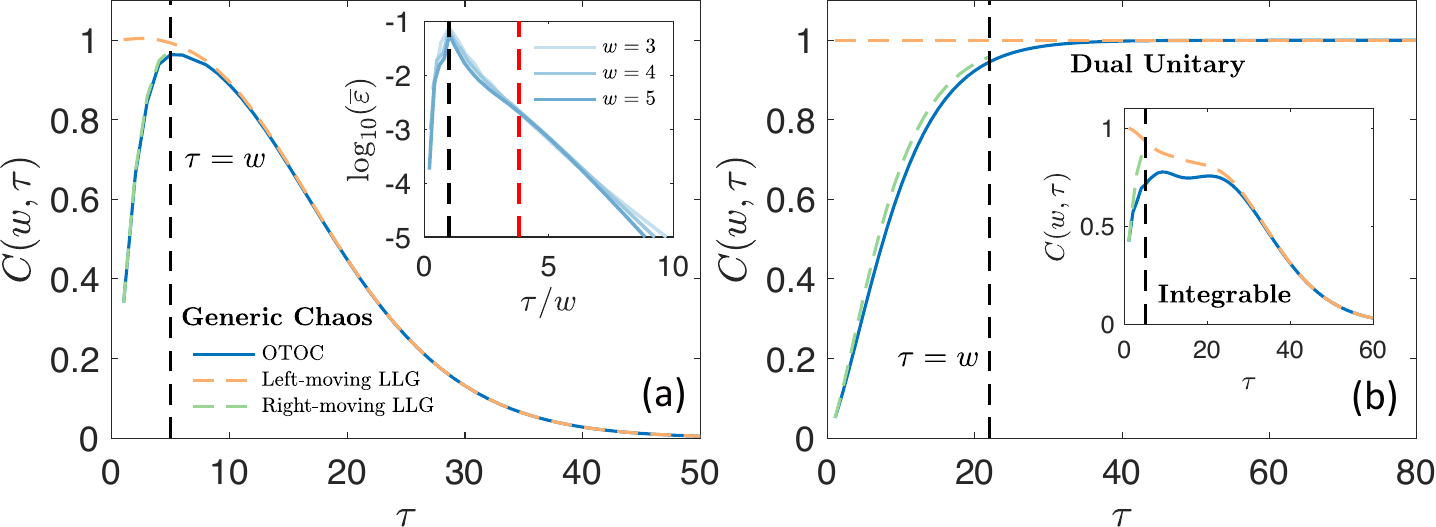}
\caption{\label{Fig:LSVA&VP} Testing the LSVA. (a) OTOC of the spatial-temporal invariant 3PM with $w=5$ for one realization. The inset shows the error of the LSVA, $\varepsilon=\abs{C-C_\text{LSVA}}$, averaged over 25 realizations, as a function of $w$. The red dashed line here is the putative position of the butterfly cone obtained in~\cite{supplementary}. (b) OTOC of the spatial-temporal random DU circuits with $w=22$, and the dashed lines represent the LSVA. The inset shows the OTOC and the LSVA for the XYZc model with $w=5$.}
\end{figure}

In contrast, for small $\tau<w/\alpha$, the leading singular value is around $q^w$, and the singular states are approximated by the two bell states $-\bra{0^w},\ket{1^w}$. 
In this region, the LSVA can be approximated by  
 \begin{align}
	C_{\text{LSVA}} \approx q^w\braket{L_w}{0^w}\braket{1^w}{R_w}=1, 
\end{align}
as shown by the orange lines in Fig.~\ref{Fig:LSVA&VP}. 
The discrepancy is due to the finite-$w$ effect. 
However, for $\tau\ll w/\alpha$, though the other singular values are exponentially smaller than the leading one, there are exponentially many of them, which cancel out the contributions to OTOC from the leading singular value. 
Nonetheless, most of the singular values decay exponentially with respect to $\tau$, while the leading one is still around $q^w$.  
Consequently, the leading singular value becomes dominant around $\tau\approx w$, resulting in the ``1'' region (and the second ``0'' region) of the OTOC. 
Particularly, we analytically and numerically show in the spatial-temporal random HRM that the leading singular value captures the behavior of the OTOC~\cite{supplementary}.
Although the leading singular value of the left-moving {\LLG}\ cannot explain the first ``0'' region $\tau\lesssim w$, it can be explained by the right-moving {\LLG}\, whose leading singular value dominates in this region.
In Fig.~\ref{Fig:LSVA&VP}(a), we study the spatial-temporal invariant 3PM, whose OTOC is well described by the LSVA. 
Further, we find that the LSVA works well for any typical realization and is more accurate in an increasingly large region in $\tau$, as shown in the inset of Fig.~\ref{Fig:LSVA&VP}(a).
However, the LSVA does not agree with the OTOC of the integrable model within the butterfly cone, as shown in Fig.~\ref{Fig:LSVA&VP}(b), but it indeed captures the behavior at the wavefront. 

\lettersection{Leading singular state}
Under Conjecture 2, we can  obtain the asymptotic behavior for the leading singular states in the large $\tau$ limit. As there is only one leading Jordan block, the leading singular states are simply given by a single pair of left and right eigenstates of that block. For {\LLG}, these eigenstates can always be generated from $F_{w=1}$ using Eq.~\eqref{Eq:Reduce}, we derive~\cite{supplementary} 
\begin{align}\label{Eq:LSS}
	\lim_{\tau\to \infty}\ket{\lambda_{w,\tau}^{\text{L}}}&\propto\ket{0^{w-1}}\Otimes{}{k}\ket{z_2^{\text{R}}},\nonumber\\
	\lim_{\tau\to \infty}\bra{\lambda_{w,\tau}^{\text{R}}}&\propto\bra{z_2^{\text{L}}}\Otimes{}{k}\bra{1^{w-1}},
\end{align}
where $\bra{z_2^{\text{L}}},\ket{z_2^{\text{R}}}$ are the left and right leading eigenstates of $F_{\text{L},w=1}$ respectively. Similar to the singular states in small $\tau$ and large $w$, the leading singular states are product states of $\left| 0\right\rangle$-s or $\left| 1\right\rangle$-s, except for one site. 
Hence, a natural ansatz for the leading singular states is $\ket{\lambda_{w,\tau}^{\text{L,VP}}}=\ket{0^{w-1}}\Otimes{}{}\ket{v_{\text{L}}(\tau)}$ and $\bra{\lambda_{w,\tau}^{\text{R,VP}}}=\bra{v_{\text{R}}(\tau)}\Otimes{}{}\bra{1^{w-1}}$,
where $\bra{v_{\text{R}}(\tau)}, \ket{v_{\text{L}}(\tau)}$ are obtained through the variational principle, giving a computational advantage over the singular value decomposition. 
In Sec. III C of \cite{supplementary}, we numerically show that the variational ansatz approximates the OTOC of the spatial-temporal invariant 3PM with good precision.

\lettersection{Dual unitarity and its perturbation}
Here we comment on the differences among the {\LLG}s for generic chaos, DU and its perturbation ~\cite{bertini_op_2020, kos2020correlations, claeys2020duotoc_vmax, bertini2020otoc,  claeys2023duotoc}.
First, the operator norms of the {\LLG}s  for DU circuits are drastically different from those of generic chaotic circuits. 
Particularly, it is analytically known that $\norm{T_w^\tau}=1$ and $\norm{F_w^\tau} = q^w$, which, along with the singular states, are $\tau$-independent.
Consequently, the OTOC cannot grow or decay exponentially along the LL direction.
Second, it can be proved that $a(z)=g(z)$ for eigenvalues $z$ with unit modulus \cite{supplementary}. 
As the leading eigenvalue of $F_w$ is $z_2(w)=1$, we have $g(z_2,w)=a(z_2,w)\geq w$.
Additionally, the {\LLG}\ of DU circuits is always block-diagonal instead of an upper-triangular one, though each block may be non-Hermitian. 
Note that the LSVA also works for DU circuits as shown in Fig.~\ref{Fig:LSVA&VP}(b), but the leading singular value behaves differently in contrast to the generic chaos case.
Lastly, the generic chaotic case has been explored by  weakly-perturbing a DU model in the so-called maximally-chaotic subspace (MCS)  \cite{claeys2023duotoc}, giving results coinciding with Eq.~\eqref{Eq:z2} and Eq. \eqref{Eq:ConjectureG} in the {\LLG} projected onto the MCS. 
However, we find that MCS are generally not invariant subspaces in generic chaotic models,
and the approximation of OTOC using MCS in spatial-temporal invariant circuits  becomes increasingly less accurate as the {\LLG} size, $w$ increases. See comparison and numerics in \cite{supplementary}.

\lettersection{Conclusion}
In this work, we study the OTOC for generic many-body quantum chaotic circuit models in terms of {\LLG}s.  
We demonstrate that the OTOC can be approximated by the leading singular value of the {\LLG}.
Using symmetries and recursivity of the {\LLG}, we propose two conjectures on the universal aspects of generic many-body quantum chaotic circuits, Eq.~\eqref{Eq:ConjectureA} and Eq.~\eqref{Eq:ConjectureG}, which are numerically tested using the large-$\tau$ decay of the OTOC.
Understanding further implications of symmetries and recursivity of the {\LLG}, studying behaviors of OTOC and  {\LLG} in the presence of local conserved quantities, and the construction of analogs of the {\LLG} in relativistic quantum field theory are natural open directions for future studies.

\lettersection{Acknowledgement} 
After the upload of this work, we are grateful for a discussion with Pieter Claeys and Michael Rampp which allowed us to clarify the similiarities and differences between \cite{claeys2023duotoc} and this work, see \cite{supplementary}.
We are grateful to John~Chalker, Andrea De Luca, Austen Lamacraft, Henning Schomerus, and Saumya Shivam for discussions and previous collaborations. 
This work is supported by the Research Grants Council of Hong Kong (Grants No.~CityU 21304720, No.~CityU 11300421, No.~CityU~11304823, and No.~C7012-21G) and City University of Hong Kong (Projects No.~9610428 and No.~7005938). 
A.C. acknowledges the support from the Royal Society grant RGS{$\backslash$}R1{$\backslash$}231444, and the fellowships from the Croucher Foundation and the PCTS at Princeton University.  
D.A.H. is supported in part by NSF QLCI grant OMA-2120757.
K.H. is also supported by the Hong Kong PhD Fellowship Scheme. 

\bibliographystyle{apsrev4-2}
\bibliography{biblio.bib}

\end{document}


\global\long\def\id{\mathbbm{1}}
\global\long\def\ui{\mathbbm{i}}
\global\long\def\ud{\mathrm{d}}

\title{Supplemental Material for ``Out-of-time-order correlator, many-body quantum chaos, light-like generators, and singular values''}

\setcounter{equation}{0} 
\setcounter{figure}{0}
\setcounter{table}{0} 
\renewcommand{\theparagraph}{\bf}
\renewcommand{\thefigure}{S\arabic{figure}}
\renewcommand{\theequation}{S\arabic{equation}}
\renewcommand\labelenumi{(\theenumi)}

\maketitle
\tableofcontents

\section{Quantum many-body models}
In this work, we focus on the quantum circuits with the brick-wall geometry as models of quantum many-body systems. Such models are defined with a evolution operator given by
\begin{align}
	U(t)=\prod_{s=1}^t\tilde U(s),\quad 
	\tilde U(s)=\bigotimes_{\substack{i \in 2 \mathbb{Z} + s\, \text{mod}\,2}}u_{i,i+1} \;.
\end{align}
For spatial-temporal invariant circuits, the two-site gates $u_{i,i+1}$ are identical for all $i$, while for spatial-temporal random circuits, $u_{i,i+1}$ are independent random variables drawn from the same ensemble. We consider 4 classes of spatial-temporal invariant or random models, as listed below. 

\subsection{Integrable model}
We consider the following many-body integrable model with $q=2$,
\begin{equation}
u_{\text{XYZc}}=\exp{i\sum_{\mu=x,y,z}a_\mu\sigma_\mu\otimes\sigma_\mu},  
\end{equation}
where $\sigma_\mu$ is the Pauli matrices, and we take  $(a_x,a_y,a_z)=(0.3,0.4,0.5)$ in the main text. The model has three symmetries: the time-reversal symmetry $\{u_{\text{XYZc}}\}^\text{T}=u_{\text{XYZc}}$, the $Z_2$ symmetry $(\sigma_z\otimes\sigma_z) u_{\text{XYZc}}  (\sigma_z\otimes\sigma_z)  = u_{\text{XYZc}}$, and the particle-hole symmetry $(\sigma_x\otimes\sigma_x) u_{\text{XYZc}} (\sigma_x\otimes\sigma_x)  = u_{\text{XYZc}}$.


\subsection{Generic chaotic models}
For generic many-body quantum chaotic models, we consider the spatial-temporal invariant or random Haar-random model (HRM)~\cite{nahum2017} and 3 parameter model (3PM)~\cite{znidaric2022},
 \begin{align}
 	u_{\text{HRM}}&=U^\text{CUE}(q^2),\\
 	u_{\text{3PM}}&= \left[U_1^\text{CUE}(2)\otimes U_2^\text{CUE}(2)\right]u_{\text{XYZc}}\nonumber\\
	&\quad \times \left[U_3^\text{CUE}(2)\otimes U_4^\text{CUE}(2)\right],
\end{align}
where $U_i^\text{CUE/COE}(n)$ are independent random matrices drawn from the circular unitary/orthogonal ensemble of degree $n$ (CUE($n$)/COE($n$)).
Here, the HRM is defined for generic $q$, while the 3PM is defined only for $q=2$.
We again take $(a_x,a_y,a_z)=(0.3,0.4,0.5)$ for the 3PM in the main text. 
We also consider the random phase model (RPM)~\cite{cdc2} for generic $q$
 \begin{align}
 	u_{\text{RPM}}&= \left[U_1^\text{CUE}(q)\otimes U_2^\text{CUE}(q)\right]\Phi\nonumber\\
	&\quad \times \left[U_3^\text{CUE}(q)\otimes U_4^\text{CUE}(q)\right],
\end{align}
where $[\Phi]_{\mu\nu}^{\mu'\nu'}=\delta_{\mu}^{\mu'}\delta_{\nu}^{\nu'}e^{i\phi_{\mu\nu}}$, and $\phi_{\mu\nu}$ are independent random variables drawn from a normal distribution with zero mean and variance $\epsilon$. In the main text, we take $q=4$ and $\epsilon=1$.
Further, we consider the $Z_2$-COE model with $q=2$ because it also has the same $Z_2$ and time-reversal symmetries as the XYZc model, given by
\begin{align}	
	&u_{Z_2\text{-COE}} \nonumber
	\\
	&=\left[\begin{array}{ccc}
	[U_1^\text{COE}(2)]_{11} & 0 & [U_1^\text{COE}(2)]_{12} \\ 
	0 & U_2^\text{COE}(2) & 0 \\ 
	{[U_1^\text{COE}(2)]_{21}} & 0 & [U_1^\text{COE}(2)]_{22}
	\end{array}\right],
 \end{align} 
where $[U]_{jk}$ denotes the $j,k$ entry of the matrix. 

\subsection{Dual unitarity (chaotic) model}
For dual unitary (DU) chaotic model~\cite{bertini_op_2020},  we consider the $q=2$ case, and the model is given by
\begin{equation}
 \begin{aligned}
 	u_{\text{DU}}&= \left[U_1^\text{CUE}(2)\otimes U_2^\text{CUE}(2)\right]u_{\text{XYZc}}\nonumber\\
	&\quad \times \left[U_3^\text{CUE}(2)\otimes U_4^\text{CUE}(2)\right]
\end{aligned}
\end{equation}
with $(a_x,a_y)=(\pi/4,\pi/4)$ and arbitrary $a_z$. In the main text, we take $a_z=0.5$.
The DU model considered here belongs to the ``completely chaotic'' DU circuits, which is the generic case within the DU family \cite{bertini_op_2020}.
Importantly, such unitary 2-site gates satisfy for following relations,
\begin{equation}
\begin{aligned}
u^{ab'}_{cd'} (u^{*})^{c' d'}_{a' b'} &= \delta_{a, a'} \delta_{c, c'}   \;,  \\
u^{a'b}_{c'd} (u^{*})^{c' d'}_{a' b'} &= \delta_{b, b'} \delta_{d, d'}   \;, 
\end{aligned}
\end{equation}
which means that unitarity also holds in the spatial direction. 

\subsection{Completely localized model}
The completely localized model for generic $q$ is taken as 
\begin{equation}
u_{\mathrm{loc}} = U^{\mathrm{CUE}}(q) \otimes U^{\mathrm{CUE}}(q)  \;,
\end{equation}
i.e., the two-site unitary gates do not couple the two sites.

\begin{figure}[b]
\centering
\includegraphics[width=\columnwidth]{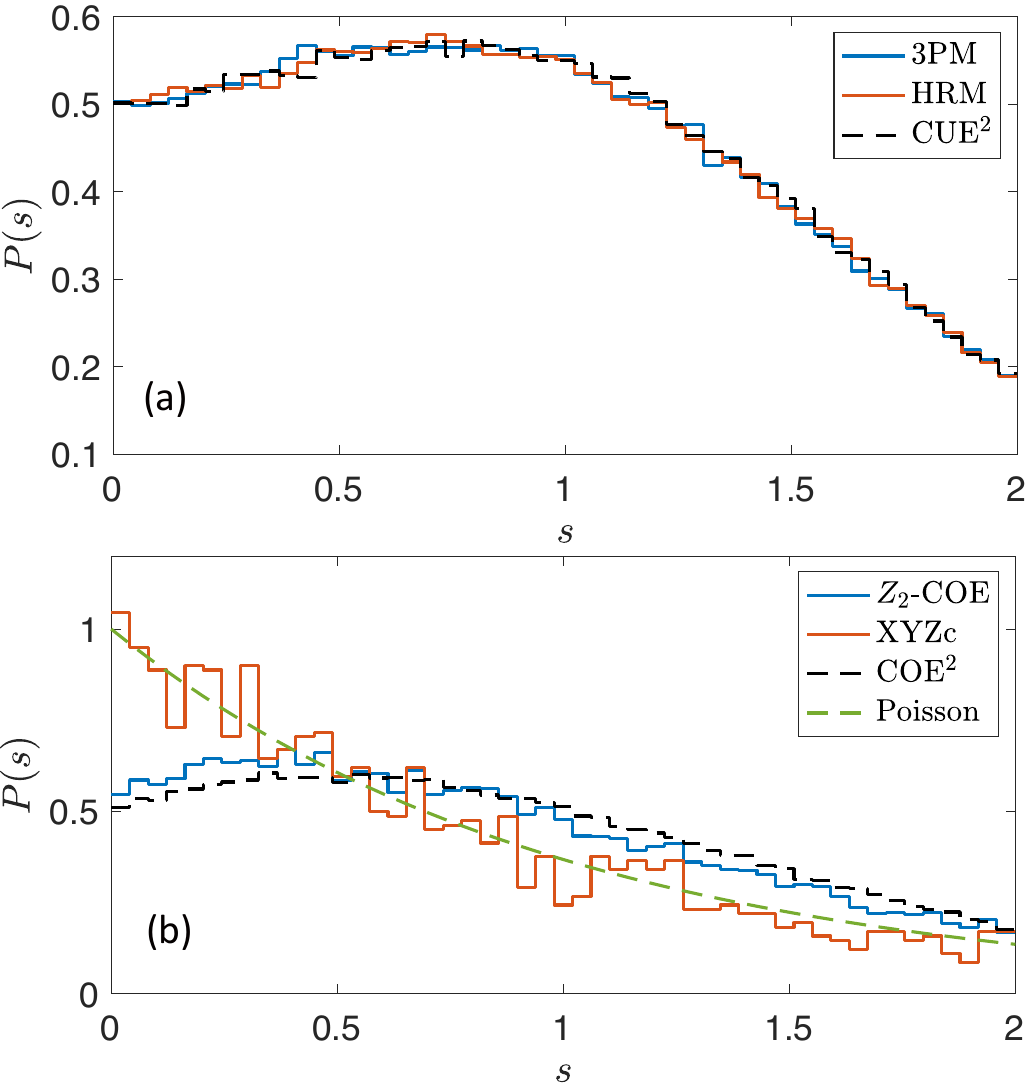}
\caption{Spacing distribution $P(s)$ of the spatial-temporal invariant models. (a) Two models without $Z_2$ and time-reversal symmetries. The system size is $L=14$, and the results are averaged over 100 realizations. (b) Two models with $Z_2$ and time-reversal symmetries. The system size is $L=16$, and the result of the $Z_2$-COE model is averaged over 10 realizations. Here, the results are in the $(2\pi/L)$-momentum sector and parity-even sector if $Z_2$ symmetry exists. What is more, we also resolve the particle-hole symmetry for the XYZc model. For the CUE$^2$ and COE$^2$ matrices, the result is averaged over 100 realizations of $2000\times2000$ matrices.
\label{Fig:CUE2}}
\end{figure}

\section{Level statistics}
Here we consider the level statistics of space-like generators or ``Floquet operators'' which generate the dynamics of the system in the time direction.
Generally, the level statistics of the (space-like) Floquet operator for a chaotic system follows the ones of Hermitian random matrix theory, like CUE or COE, whereas it may manifest other level statistics in the presence of space-time translational symmetry. 
We say that a Floquet system has space-time translational (STT) symmetry if its evolution operator $U(t_2,t_1)$ satisfies
\begin{align}\label{Eq:SFF}
    \mathcal{T}U(t_2,t_1)\mathcal{T}^{-1}=U(t_2+T/n,t_1+T/n),
\end{align}
where $\mathcal T$ is the translation by one site, $T$ is the period of the Floquet system, and $n$ is a positive integer. Note that $n=2$ for the brick-wall structure studied in this work. Thus, the STT symmetry spontaneously implies that 
\begin{align}
    \mathcal{T}^nU(t_2,t_1)\mathcal{T}^{-n}=U(t_2+T,t_1+T)=U(t_2,t_1),
\end{align}
that is, the system is invariant under translation by $n$ sites. Additionally, the Floquet operator in this system is reduced to
\begin{align}
    F&=U(T,T-T/n)\times\cdots\times U(T/n,0)\nonumber\\
    &=\mathcal{T}^{(n-1)}U_0\mathcal{T}^{-(n-1)}\times \mathcal{T}^{(n-2)}U_0\mathcal{T}^{-(n-2)}\times \cdots \times U_0\nonumber\\
    &=\mathcal{T}^{n}(\mathcal{T}^{-1}U_0)^n=(\mathcal{T}^{-1}U_0)^n\mathcal{T}^{n},
\end{align}
where $U_0=U(T/n,0)$
In a given momentum subspace, $\mathcal{T}^n$ is constant, and therefore, we only need to consider
\begin{align}
    F'&=(\mathcal{T}^{-1}U_0)^n.
\end{align}
If $F'$ is chaotic, then $\mathcal{T}^{-1}U_0$ should not be integrable because the power of an integrable operator is still integrable. However, if $\mathcal{T}^{-1}U_0$ is chaotic and we assume it to follow CUE or COE, $F'$ cannot be described by CUE or COE. Instead, $F'$ should be $\text{CUE}^n$ or $\text{COE}^n$ which is defined as
\begin{align}
    \text{CUE}^n:=\{S^n:S\in\text{CUE}\},\nonumber\\
    \text{COE}^n:=\{S^n:S\in\text{COE}\}.
\end{align}
Particularly, we verify the CUE$^2$ and COE$^2$ by studying the level statistics in the spatial-temporal invariant 3PM, HRM, and $Z_2$-COE model as shown in Fig.~\ref{Fig:CUE2}. For the XYZc model, the level statistics indicates that it is integrable.

\section{Reducibility and consequences}
\subsection{Algebraic multiplicity}
\begin{figure}[t]
\center
\includegraphics[width=0.8\columnwidth]{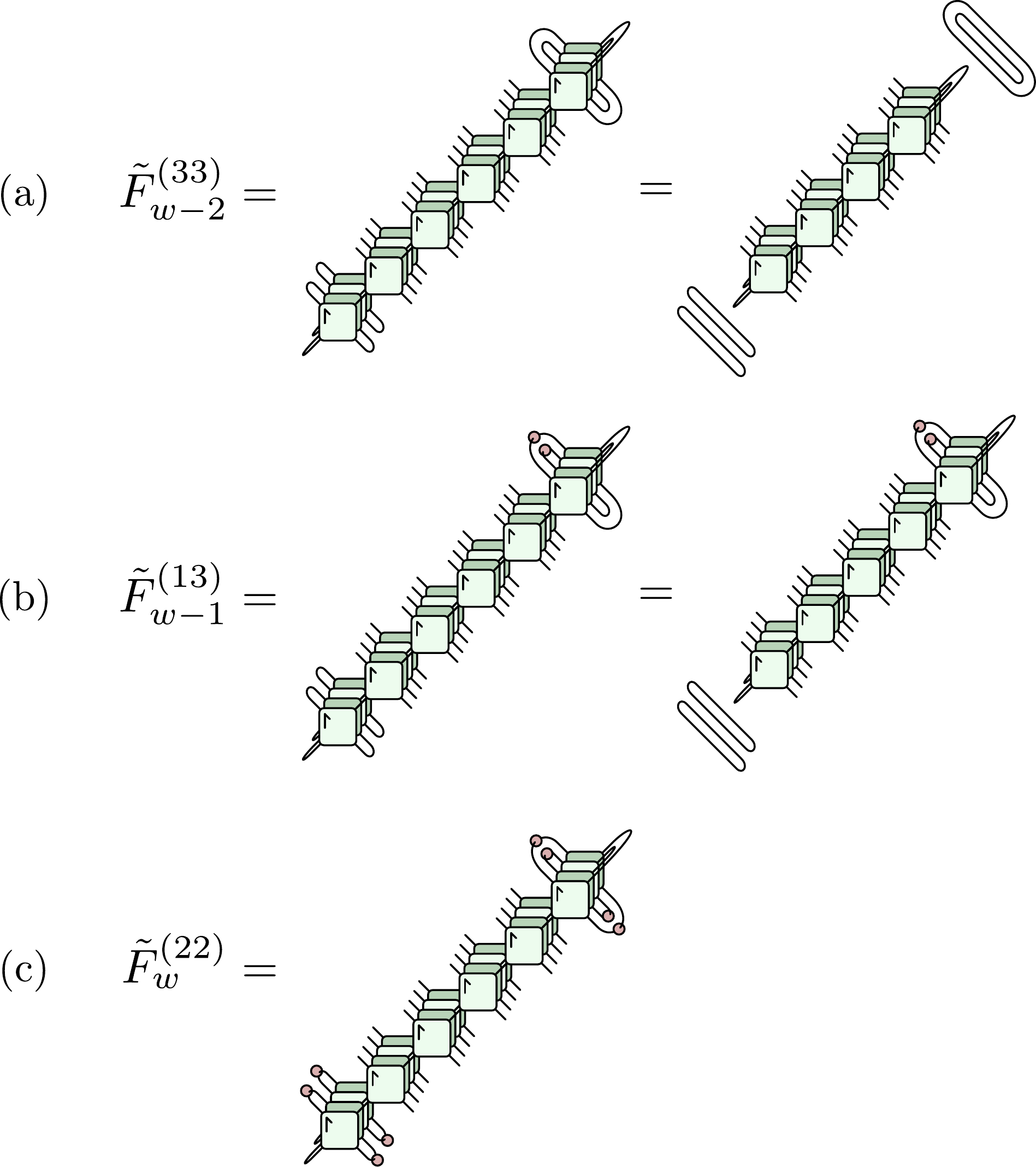}
\caption{\label{fig:block_eg} Three example blocks in Eq.~\eqref{eq:block_mat}, where red dots represent operators that are orthogonal to the identity operator.}
\end{figure}

Nondiagonalizability appears only if there are degeneracies (algebraic multiplicity greater than $1$). In this section, we show that the degeneracies result from the reducibility introduced in the main text. To start, we consider the block-upper-diagonal structure ($w\geq 2$) shown in the main text,
\begin{align}\label{eq:block_mat}
	F_w=\left[\begin{array}{cccc}
	\tilde{F}_{w-1}^{(11)} & \tilde{F}_w^{(12)} & \tilde{F}_{w-1}^{(13)} & \tilde{F}_{w}^{(14)} \\
	 0 & \tilde{F}_w^{(22)} & 0  & \tilde{F}_w^{(24)} \\
	  0  & 0  &  \tilde{F}_{w-2}^{(33)} & \tilde{F}_{w-1}^{(34)} \\
	 0 & 0  &  0 & \tilde{F}_{w-1}^{(44)}\end{array}\right].
\end{align}
See Fig.~\ref{fig:block_eg} for the diagrammatical representation of a few blocks as examples.
It is easy to see that for $w>2$, $\tilde F_{w-2}^{(33)}$ is isomorphic to $F_{w-2}$, and therefore they have the same eigenvalues. For $w=2$, $\tilde F_{w-2}^{(33)}$ is trivially zero. Similarly, there is another isomorphism,
\begin{align}
	F_{w-1}\cong\left[\begin{array}{cc} \tilde F_{w-1}^{(11)} & \tilde F_{w-1}^{(13)} \\
	0 & \tilde F_{w-2}^{(33)}
	\end{array}\right]\cong
	\left[\begin{array}{cc}\tilde F_{w-2}^{(33)} & \tilde F_{w-1}^{(34)} \\
	0 & \tilde F_{w-1}^{(44)}
	\end{array}\right].
\end{align}
Hence, the algebraic multiplicity of $z\neq 0$ in $\tilde F_{w-1}^{(11)},\tilde F_{w-1}^{(44)}$ is $a(z,w-1)-a(z,w-2)$, and adding up four diagonal blocks, we obtain the recursive relation for $w\geq2$,
\begin{align}\label{Eq:Recursive}
	a(z,w)=2a(z,w-1)-a(z,w-2)+\tilde a(z,w).
\end{align}

Next, using Eq.~\eqref{Eq:Recursive}, we will prove that if $z$ is an eigenvalue of $w$, then $a(z,w+1)\geq a(z,w)+1$.
Suppose $z$ is an eigenvalue of $F_{w_0}$ but not $F_{w_0-1}$ for some $w_0$.
If $w_0=1$, then we have
\begin{align}\label{eq:recur}
	a(z,w_0+1)&=2a(z,w_0)+\tilde a(z,w_0+1)\nonumber\\
	&\geq a(z,w_0)+1,
\end{align}
where we use the fact that $\tilde F_{0}^{(33)}=0$. If $w_0>1$, then we have $a(z,w_0-1)=0$ and we have also Eq.~\eqref{eq:recur}.
Hence, we have
\begin{align}
	a(z,w)-a(z,w-1)&\geq a(z,w-1)-a(z,w-2)\nonumber\\
	&\geq a(z,w_0+1)-a(z,w_0)\geq1.
\end{align}
Note that the equality holds if $\tilde a(z,w)=0$ for all $w>w_0$.

\subsection{Irreducibility of the leading singular state}

In Section~\ref{Sec:Eigen}, we prove that the OTOC possesses the following scaling
\begin{align}
\limsup_{\tau\to\infty}\frac{C(w,\tau)}{\tau^{\varphi[z_2(w),w]}\abs{z_2(w)}^\tau}<+\infty,
\end{align}
where $\varphi(z,w)=N_{\max}(z,w)-1$, and $N_{\max}(z,w)$ is the dimension of the largest Jordan block corresponding to eigenvalue $z$. 
What is more, the limit is nonvanishing if the initial and final states have nonvanishing overlap with the leading singular states.
If there only one Jordan block corresponding to $z_2(w)$ for all $w$, then we have $\varphi[z_2(w),w]=a[z_2(w),w]-1$ and the scaling is greater for larger $w$, because either the algebraic multiplicity of $z_2(w-1)$ increases or $z_2(w)$ is greater than $z_2(w-1)$.
Hence, we have $C(w,\tau)\gg C(w-1,\tau)$ in the large $\tau$ limit unless the initial and final states of $C(w,\tau)$ have nonvanishing overlap with the leading singular states. 
Consider an arbitrary pair of initial and final states $\ket{v_\text{R}},\bra{v_\text{L}}$ of width $(w-1)$, and we have
\begin{align}
C(w-1,\tau)&=\mel{v_\text{L}}{F_{w-1}^\tau}{v_\text{R}}\nonumber\\
&=\mel{v_\text{L}'}{F_{w}^\tau}{v_\text{R}'}=C'(w,\tau),
\end{align}
where
\begin{align}
\bra{v_\text{L}'}=\bra{v_\text{L}}\Otimes{}{}\bra{1},\quad
\ket{v_\text{R}'}=\ket{v_\text{R}}\Otimes{}{}\ket{v}.
\end{align}
Hence, $\ket{v}$ is an arbitrary state satisfying $\braket{1}{v}=1$. 
If $\ket{v_\text{R}'}, \bra{v_\text{L}'}$ have nonvanishing overlap with the lead singular states, then $C'(w,\tau)\gg C(w,\tau)$ for large enough $\tau$, which is impossible. As $\ket{v_\text{L}}$ and $\ket{v_\text{R}}$ are both arbitrary, the only way to have vanishing overlap is that $(\bra{v_\text{L}}\Otimes{}{r}\bra{1})\ket{\lambda_\text{L}}=0$ for all $\bra{v_\text{L}}$, which is just the definition of ``irreducibility for bra states''. Similarly, we can show that $\ket{\lambda_\text{R}}$ is also irreducible.

For instance, if we further assume $z_2(w)=z_2(w=1)$, utilizing Eq.~\eqref{Eq:SingularState}, we have
\begin{align}\label{Eq:LSS}
	\bra{\lambda^\text{L}_{\tau=\infty}}&\propto\bra{z_{2}^{\text{R}}(w)}=\bra{0^{w-1}}\Otimes{}{r}\bra{z_2^{\text{R}}(w=1)},\nonumber\\
	\ket{\lambda^\text{R}_{\tau=\infty}}&\propto\ket{z_2^{\text{L}}(w)}=\ket{z_2^{\text{L}}(w=1)}\Otimes{}{r}\ket{1^{w-1}},
\end{align}
where $\bra{z_2^{\text{L}}(w)},\ket{z_2^{\text{R}}(w)}$ are the left and right eigenstate of eigenvalue $z_2(w)$ respectively. Incorporating Eq.~\eqref{Eq:SingularState}, we can readily verify the irreducibility because $\braket{1}{z_2^{\text{R}}(w=1)}=\braket{0}{z_2^{\text{L}}(w=1)}=0$.

\subsection{A variational method for leading singular state}
\begin{figure}[t]
\center
\includegraphics[width=\columnwidth]{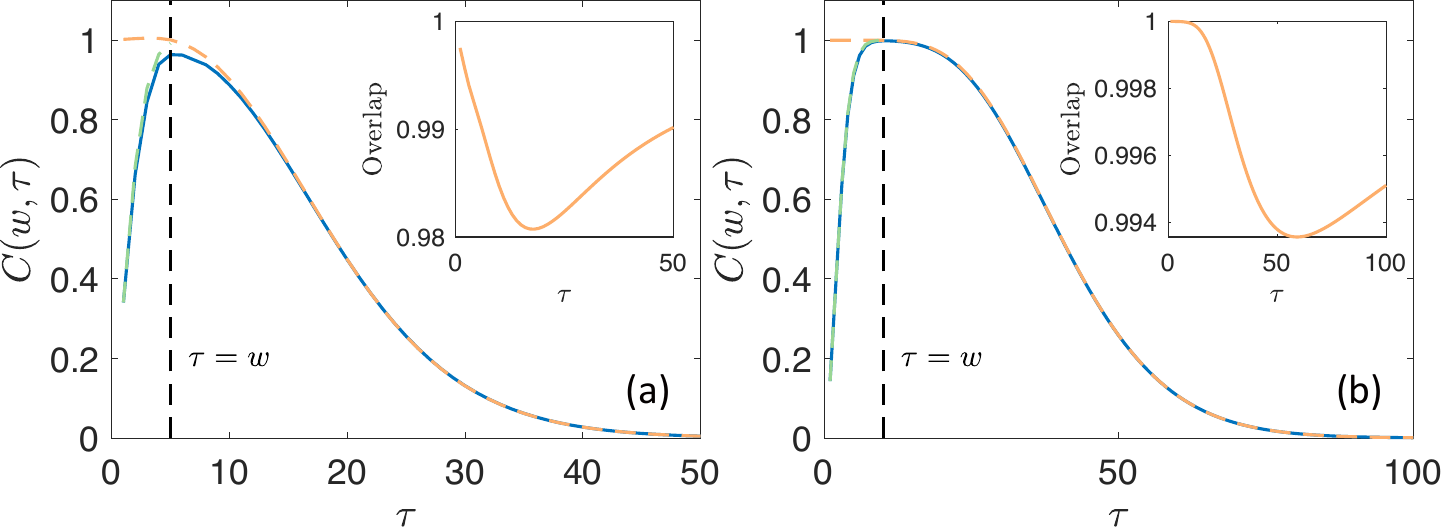}
\caption{\label{Fig:VP}OTOC in (a) spatial-temporal invariant 3PM with $w=5$ for one realization, (b) spatial-temporal random HRM with $w=10$ and $q=2$. Here, the blue lines represent the exact results, the orange lines use the left-moving LL generator, and the green lines use the right-moving one. The inset shows the overlap between the exact leading singular state and the variational ansatz $\abs{\braket{\lambda_{w,\tau}^{\text{L,VP}}}{\lambda_{w,\tau}^\text{L}}\braket{\lambda_{w,\tau}^{\text{R}}}{\lambda_{w,\tau}^\text{R,VP}}}$.}
\end{figure}

In the main text, we devise the following variational ansatz for the leading singular states,
\begin{align}
 &\ket{\lambda_{w,\tau}^{\text{L,VP}}}=\ket{0^{w-1}}\Otimes{}{}\ket{v_{\text{L}}(\tau)}\nonumber\\
 &\bra{\lambda_{w,\tau}^{\text{R,VP}}}=\bra{v_{\text{R}}(\tau)}\Otimes{}{}\bra{1^{w-1}},
\end{align}
where $\bra{v_{\text{R}}(\tau)}, \ket{v_{\text{L}}(\tau)}$ are obtained variationally.
In Fig.~\ref{Fig:VP}, we numerically show that the variational ansatz approximates the OTOC of the spatial-temporal invariant 3PM and the spatial-temporal random HRM with good precision. The high overlap between the variational state and the exact singular state also indicates that the ansatz indeed captures the primary structure of the leading singular states in generic models.

\section{Asymptotic decay rate}

\begin{figure}[b]
\center
\includegraphics[width=\columnwidth]{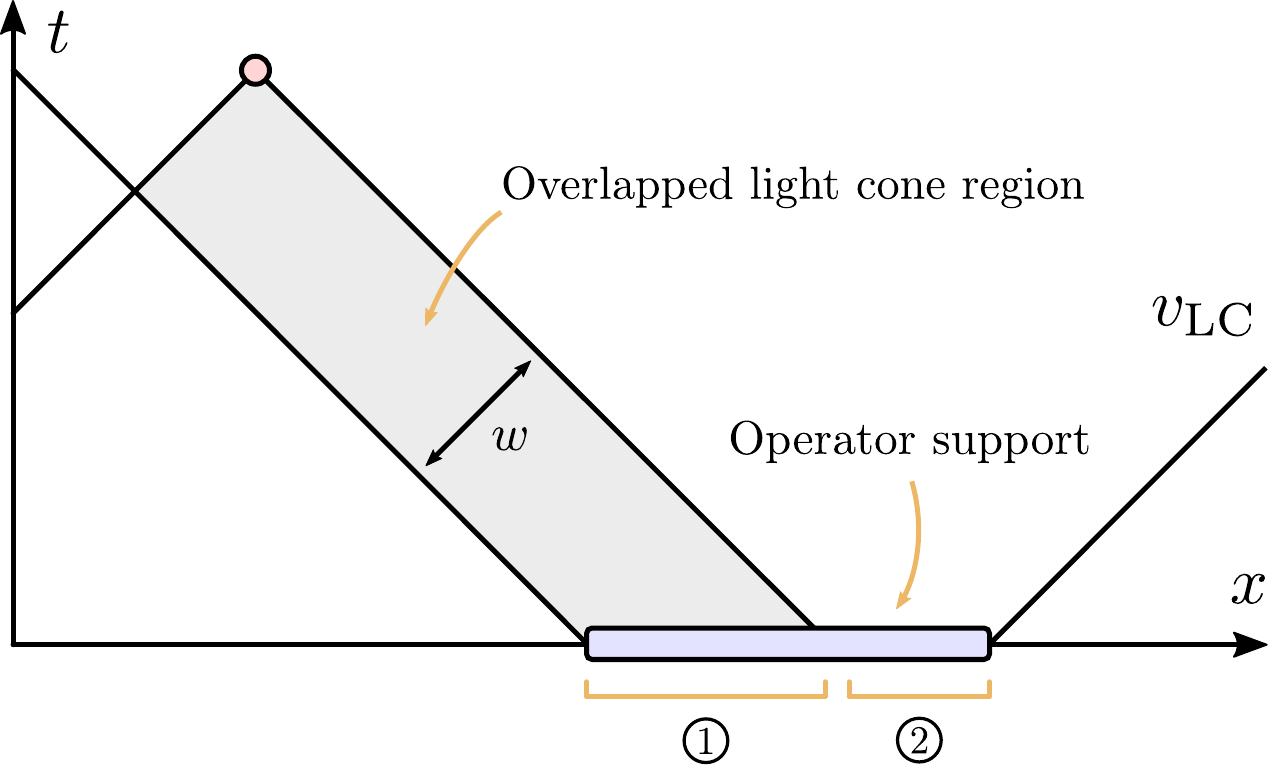}
\caption{\label{Fig:z2_proof} OTOC of operator $A$ (blue) with large operator support and $B$ (red) with local operator support only depend on the overlapped light cone region with width $w$.}
\end{figure}

According to Section~\ref{Sec:Eigen}, we know that the decay rate of the OTOC along LL direction is $z_2(w)+\varphi(w)\ln(\tau)/\tau$, which becomes $z_2(w)$ in the large $\tau$ limit. As illustrated in Fig.~\ref{Fig:z2_proof}, the operator $B$ moves along the LL direction, and the decay rate is thus $z_2(w)$. Particularly, the decay rate is $z_2(w=1)$ if $B$ is exactly on the lightcone of the operator $A(t)$. From the reducibility of the \LLG, we have $z_2(w)\geq z_2(w=1)$, so the decay rate is not less than $z_2(w=1)$ for $B$ inside the lightcone. Nonetheless, we will show that the $z_2(w=1)$ is asymptotically the maximal decay rate of the system, meaning that $z_2(w)=z_2(w=1)$.

We can decompose $A(t)$ into Pauli strings,
 \begin{align}
	A(t)=\sum_{x=-t}^{+\infty}A_x(t),
\end{align}
where $A_x(t)$ denotes the component of $A(t)$ that is supported on $[x,+\infty]$ and not identity on site $x$. It is obvious that only $A_x(t)$ with $x\in[-\tau,w'-\tau]$ contributes to $C(w',\tau)$, and therefore $C(w, \tau)$ and $C(w-1,\tau)$ can be estimated by
\begin{align}
	C(w-1,\tau)&\sim\sum_{x=-\tau}^{w-\tau-1}\norm{A_x(t)}_\text{F}^2,\nonumber\\
	C(w,\tau)&\sim \norm{A_{w-\tau}(t)}_\text{F}^2+\sum_{x=-\tau}^{w-\tau-1}\norm{A_x(t)}_\text{F}^2
\end{align}
If $z_2(w)>z_2(w-1)$ for some $w$, then $C(w,\tau)\gg C(w',\tau)$ in the large $\tau$ limit for all $w'<w$. It means that 
\begin{align}\label{Eq:PauliString}
	\norm{A_{w-\tau}(t)}_\text{F}^2\gg\sum_{x=-\tau}^{w-\tau-1}\norm{A_x(t)}_\text{F}^2,
\end{align}
and that the OTOC between $A(t)$ and $B$ is dominated by the OTOC between $A_{w-\tau}(t)$ and $B$. However, as $B$ is exactly on the lightcone of $A_{w-\tau}(t)$, the asymptotic decay rate of the OTOC between $A(t)$ and $B$ cannot be greater than $z_2(w=1)$, leading to a contradiction. Hence, $z_2(w)$ should equal $z_2(w-1)$ instead. In Section~\ref{Sec:z2_proof}, this argument is rigorously proved under a mild assumption.

Note that Eq.~\eqref{Eq:PauliString} is true as long as $C(w,\tau)$ has a greater scaling than $C(w-1,\tau)$. Therefore, if $z_2(w)=z_2(w-1)$ but $\varphi[z_2(w),w]>\varphi[z_2(w-1),w-1]$, Eq.~\eqref{Eq:PauliString} also holds true. Notwithstanding, there is no contradiction because $\varphi[z_2(w),w]$ does not affect the asymptotic decay rate. For generic chaotic models, our conjecture implies an increase of $\varphi[z_2(w),w]$, and $\norm{A_{w-\tau}(t)}_\text{F}^2/\norm{A_{w-\tau-1}(t)}_\text{F}^2$ should scale linearly with respect to $t$, instead of an exponential of $t$.

\section{Spatial-temporal random HRM}
\subsection{Eigenvalues and multiplicity}
For the spatial-temporal random circuit, the ensemble-averaged \LLG\ is given by
\begin{align}
	T_w=\overline{\rbra{0}\Bigodot{w}{r}\UU_2\rket{1}}=\rbra{0}\Bigodot{w}{r}\overline{\UU_2}\rket{1},
\end{align}
where the overline denotes the ensemble average. For the second equality, we use the fact that $\UU_2$s are independent random variables. In this section, we study the spatial-temporal random HRM, whose $\overline{\UU_2}$ is given by
\begin{align}\label{Eq:Original Tensor}
	\overline{\UU_2}=\tilde M_{j_1 j_2}^{i_1i_2}\ket{i_1}\rket{j_1}\rbra{i_2}\bra{j_2},
\end{align}
where $\tilde M_{j_1 j_2}^{i_1i_2}=\delta^{i_1i_2}\delta_{j_1j_2}q^2\text{Wg}(i_1,j_1,q^2)$, and 
\begin{align}
	\Wg(i,j, N) =
	\begin{cases}
	(N^2-1)^{-1}, &i=j\\
	-[N(N^2-1)]^{-1},&i\neq j.
	\end{cases}
\end{align}
We prove in Section~\ref{Sec:Case A HRM} that the eigenvalues of the \LLG\ are
\begin{align}
\varepsilon_n=
\begin{cases}
	\left(\frac{q}{q^2+1}\right)^n & n \text{\ even},\\
	q\left(\frac{q}{q^2+1}\right)^n& n \text{\ odd},
\end{cases}
\end{align}
with algebraic multiplicity $w\choose n$. Here, $n$ takes value from $0,1,\cdots,w-1$. Further, the geometric multiplicity of the subleading eigenvalue $z_2=q^2/(q^2+1)$ is one.

\subsection{Extracting butterfly velocity from leading singular value}

\begin{figure}[t]
\center
\includegraphics[width=\columnwidth]{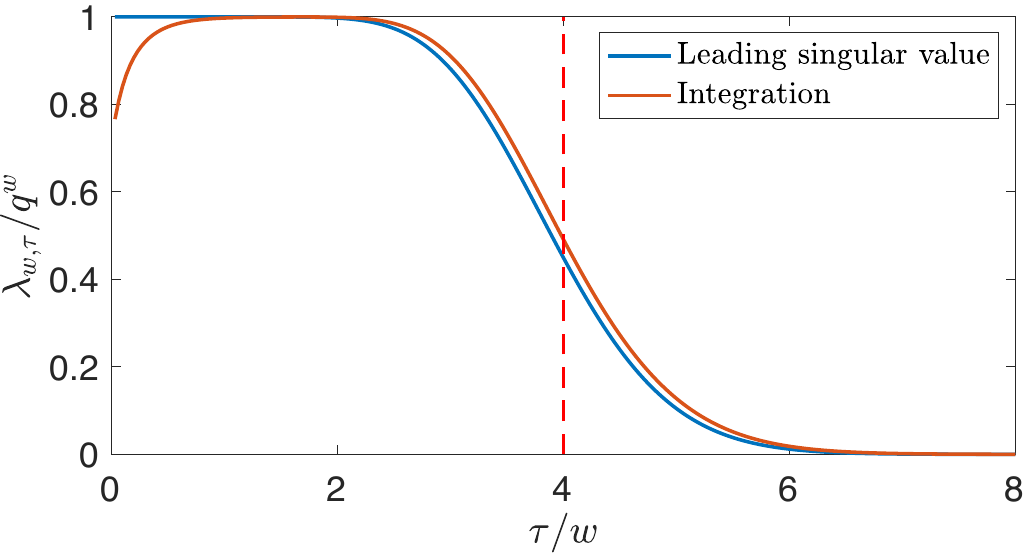}
\caption{\label{Fig:Integral} Leading singular value of $F_w$ in the spatial-temporal random HRM and the approximation Eq.~\eqref{Eq:Integral}. The red dashed line denotes the position of the butterfly cone $\tau/w=q^2$. Here, we take $w=30$ and $q=2$.}
\end{figure}

In Section~\ref{Sec:Case A HRM}, using Stirling's formula, we also show that the leading singular value of $F_w$ around the butterfly cone is approximated by
 \begin{align}\label{Eq:Integral}
 	\lambda_{w,\tau}\approx q^w f_\tau(w/\tau),
\end{align}
where $f_\tau(x)$ is defined by
\begin{align}
	f_\tau(x)&:=\int_{0}^{x}\dd s\sqrt{\frac{z_2\tau}{2\pi(1-z_2)(1+s)}}\nonumber\\
	&\ \ \times\exp \qty{\frac{[(1-z_2)s-z_2]^2}{2z_2(1-z_2)(1+s)}\tau}\nonumber\\
	&=\int_{0}^{x\tau}\dd w'\sqrt{\frac{z_2}{2\pi(1-z_2)(w'+\tau)}}\nonumber\\
	&\ \ \times\exp \qty{\frac{[(1-z_2)w'-z_2\tau]^2}{2z_2(1-z_2)(\tau+w')}}.
\end{align}
The approximation is numerically verified in Fig.~\ref{Fig:Integral} to be able to capture the leading singular value around the butterfly cone.
It is noteworthy to mention that $f_\tau(x)$ is actually an integral of a Gaussian function along the LL direction. To see it, we transform the coordinates back to $(x,t)$ though $x=w'-\tau$ and $t=w'+\tau$, and the integrand becomes
\begin{align}
	&\sqrt{\frac{z_2}{2\pi(1-z_2)t}}\exp \qty{\frac{[x+(1-2z_2)t]^2}{8z_2(1-z_2)t}}\nonumber\\
	=&\sqrt{\frac{z_2}{2\pi(1-z_2)}}\frac1{\sqrt{t}}\exp \qty{\frac{[x+v_bt]^2}{2\sigma(t)^2}}
\end{align}
with $v_b=(q^2-1)/(q^2+1)$ and $\sigma(t)=2 q\sqrt{t}/(q^2+1)$. Here, we recover the butterfly velocity and the diffusion in this model by analyzing the behavior of the leading singular value.

\section{Light-like generators for two special cases}

\subsection{Dual unitary models}

For DU circuits, we have $T_w\ket{1^w}=T_w^\dag\ket{1^w}=\ket{1^w}$ and $T_w\ket{0^w}=T_w^\dag\ket{0^w}=\ket{0^w}$, implying that the subspace spanned by $\ket{1^w},\ket{0^w}$ is an invariant subspace of both $T_w$ and $T_w^\dag$. Hence, we only need to consider the restriction of the \LLG\ on this subspace, and we have
\begin{align}
	T_w^\tau=\left[\begin{array}{cc}  1&   \\  &1  \end{array}\right],\quad
	F_w^\tau=\left[\begin{array}{cc}  0&  -\sqrt{q^{2w}-1} \\0  &1  \end{array}\right],
\end{align}
where we define $\ket{0^w}=[q^{w/2},0]^\top$ and $q^{w/2}\ket{1^w}=[1,\sqrt{q^{2w}-1}]^\top$.
Hence, $\norm{F_w^\tau}=q^w$.

\begin{figure}[H]
\center
\includegraphics[width=\columnwidth]{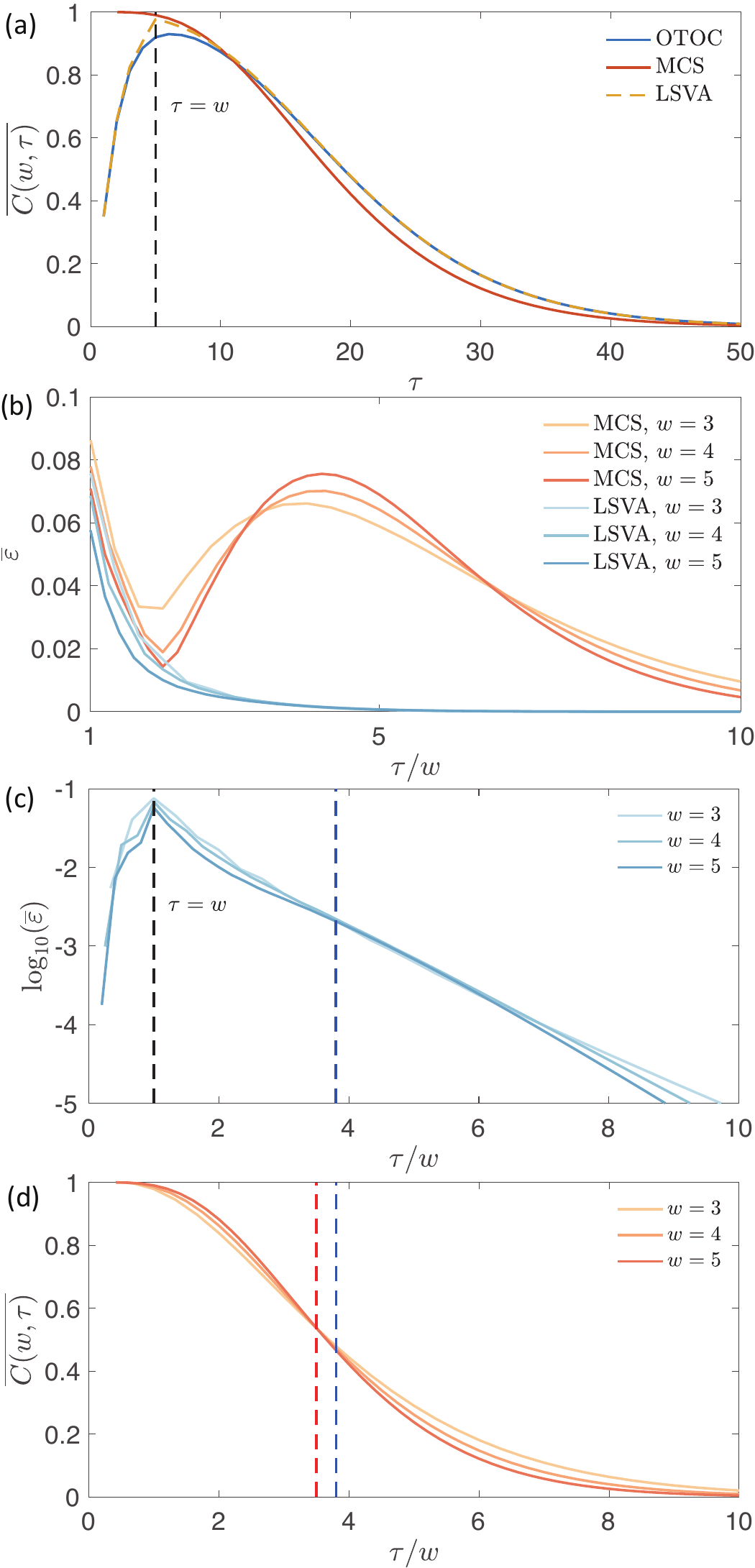}
\caption{\label{Fig:MCS} (a) The value of OTOC, the OTOC projected on the MCS, and the LSVA of the spatial-temporal invariant 3PM with $w=5$. Here, the results are averaged over the same 25 realizations as those in Fig.~\ref{Fig:3PM_EnsembleAve}. (b) Ensemble average of the error $\varepsilon=\abs{C-C_{\text{MCS}}}$, and $\varepsilon=\abs{C-C_{\text{LSVA}}}$. (c) Error of the LSVA in the log scale. (d) OTOC projected on the MCS for different $w$. The red dashed line denotes the cross point of the OTOC projected on the MCS, and marks the effective butterfly velocity associated with this method. The blue dashed lines denote the position of corresponding butterfly velocity, $v_{\mathrm{b}}$, at $\tau/w \approx 3.8$, which is extracted in an analogous figure for LSVA given by Fig.~\ref{Fig:3PM_EnsembleAve}. 
}
\end{figure}

\subsection{Completely localized models}

If the two-site gate is a tensor product of two one-site gates, we then call the circuit a completely localized circuit. 
The {\LLG}s of the completely localized circuits satisfy
\begin{align}
	T_w\ket{0^m}\otimes\ket{1^n}=q\ket{0^{m+1}}\otimes\ket{1^{n-1}},\nonumber\\
	T_w^\dag\ket{0^m}\otimes\ket{1^n}=q\ket{0^{m-1}}\otimes\ket{1^{n+1}},
\end{align}
where $m+n=w$ and $m,n>0$. This implies $\norm{T_w}\geq q$, which in conjunction with Eq.~\eqref{Eq:UpperBound} results in $\norm{T_w}=q$. Moreover, the subspace spanned by all $\ket{0^m}\otimes\ket{1^{w-m}}$ with $m=0,1,\cdots,w$ is an invariant subspace of both $T_w$ and $T_w^\dag$. Within this subspace, we particularly choose the following orthonormal basis, $\ket{e_w}=\ket{0^w}/q^{w/2}$ and for $m=0,1,\cdots, w-1$
\begin{align}
	\ket{e_m}=\ket{0^{m}}\otimes\frac{q\ket{1}-\ket{0}}{q^{w/2}\sqrt{q^2-1}}\otimes\ket{1^{w-m-1}}.
\end{align}
Under this basis, for $\tau<w$, $T_w^\tau$ becomes
\begin{align}
	T_w^\tau=\left[\begin{array}{ccccccc} 
	0 &   &   &   &  && \\ 
	\vdots &   &   &   &  && \\ 
	0 &   &   &   &  && \\ 
	q^\tau &  0 &   &   &   &&\\ 
	 &  \ddots &  \ddots &  &   &&\\
	   &   &  q^\tau & 0  &  \ldots  & 0 & \\
	     &   &  & \chi(\tau) & \ldots & \chi(1) & 1\end{array}\right],
\end{align}
where $\chi(\tau)=q^{\tau-1}\sqrt{q^2-1}$. Because $q^{w/2}\ket{1^w}=[\chi(w),\chi(w-1),\cdots,\chi(1),1]^\top$,
we obtain $\norm{F_w^\tau}=q^w$ for $\tau<w$ and $\norm{F_w^\tau}=0$ for $\tau\geq w$.

\section{Difference between this work and perturbative results around DU limt in \cite{claeys2023duotoc}}

Here we comment on the differences between the approach used in \cite{claeys2023duotoc} and this paper.
Consider the DU maximally chaotic subspace (MCS) defined as
\begin{equation}
\begin{aligned}
\mathrm{MCS} = & \; \mathrm{Span} \{ \ket{m, w- m}= 	\ket{0^m}\otimes\ket{1^{w-m}} ,  
\\
&  \qquad \qquad \qquad 
 m= 0, 1, 2, \dots, w  \} \;.
	\end{aligned}
\end{equation}
 \cite{claeys2023duotoc} approximated the OTOC by projecting the {\LLG} onto the MCS for a model weakly perturbed from its dual unitarity (DU) limit, and derived (i) $v_{\mathrm{b}} < v_{\mathrm{LC}}$ [For DU circuits, it has been found that $v_{\mathrm{b}} = v_{\mathrm{LC}}$, see \cite{kos2020correlations, claeys2020duotoc_vmax}], and (ii) a diffusive front of the OTOC, coinciding with the one found in generic chaotic systems \cite{nahum2018, keyserlingk2018}.  

In relation to this work, for the {\LLG} projected onto the DU MCS in the weakly perturbed DU circuit, \cite{claeys2023duotoc} showed that (i) the algebraic and geometric multiplicity of the 
subleading eigenvalue, $z_2$, of the {\LLG} projected onto the MCS is  $w$ and $1$ respectively;
and  that (ii) the value of $z_2$ is independent of $w$. These latter two statements coincide with the statements in the main text postulated and derived on $z_2$ in the generic chaotic case.
However, we find that the projection of {\LLG} onto the MCS of generic chaotic circuits does not capture OTOC in general, due to the following reasons:

\begin{enumerate}
\item[1.] \textbf{The MCS is generally not an approximately invariant subspace in spatial-temporal invariant generic chaotic circuits.} 
As the MCS is an exact invariant subspace for DU circuits, one might assume that it also works for weakly-perturbed DU circuits. However, the MCS is generally far from invariant in spatial-temporal invariant generic chaotic circuits.
For example, this can be understood by considering \LLG s of width $w=1$. For the MCS to perform well, the subleading eigenvector of $T_{w=1}$ must be approximately in the MCS. However, the MCS of $w=1$ is a 2-dimensional subspace of a $q^4$-dimension Hilbert space, and the subleading eigenvector is generally far from such a small subspace. However, we numerically observe that the MCS works generically well in the averaged OTOC dynamics of spatial-temporal random circuits. We surmise that this is because the $q^4$-dimension Hilbert space is reduced to a 2-dimensional subspace after ensemble averaging. Therefore, the MCS of $w=1$ becomes the whole Hilbert space of $T_{w=1}$.
\item[2.] \textbf{Accuracy of OTOC calculated from {\LLG} projected onto MCS decreases as $w$ increases in spatial-temporal invariant generic chaotic circuits.} 
Numerically, we test the OTOC projected onto MCS in Fig.~\ref{Fig:MCS}(a-c) for the spatial-temporal invariant 3PM. In Fig.~\ref{Fig:MCS}(a), we find that there is a notable discrepancy between the OTOC and the projected OTOC. Furthermore, the accuracy of the projected OTOC decreases with increasing $w$, while the LSVA performs better for increasing $w$ as shown in Fig.~\ref{Fig:MCS}(b) and~\ref{Fig:MCS}(c).
\item[3.] \textbf{The value of the butterfly velocity $v_{\mathrm{b}}$ is not captured by the {\LLG} projected onto MCS.}
In Ref.~\cite{claeys2023duotoc}, the error of the butterfly velocity comes from two approximations. The first one is the first-order perturbation that projects the \LLG\ onto the MCS. The second one is the path integral method that treats the off-diagonal terms in the projected \LLG\ perturbatively. In principle, the second error can be calculated order by order, and Ref.~\cite{claeys2023duotoc} obtained the analytical result up to the second order. However, the first error cannot be readily corrected by higher orders, and we numerically observe this error in Fig.~\ref{Fig:MCS}(a) and~\ref{Fig:MCS}(d).
\end{enumerate}

The averaged OTOC dynamics of the spatial-temporal random HRM is exactly solvable \cite{nahum2018, keyserlingk2018}.
For such spatial-temporal random circuits, the OTOC can be written in terms of a (spatial-temporal invariant) {\LLG} after performing the ensemble average in space and time.
In \ref{Sec:Case A HRM}, we derive the eigenvalues, multiplicities, leading singular values, and the butterfly velocity associated to the {\LLG}.
As a side result from the analysis in \ref{Sec:Case A HRM}, for the LLG of the averaged OTOC dynamics of the spatial-temporal random HRM, we derive that the MCS is an exact invariant subspace, and consequently, the {\LLG} projected onto MCS can capture the OTOC in this model. 
However we expect that for generic spatial-temporal circuits, the MCS is not exactly invariant, because unlike DU circuits, in generic chaotic circuits, the invariance subspaces of the {\LLG}, $T_w$, and its hermitian conjugate, $T^\dagger_w$, are generally not the same. We numerically verify that the MCS is not an exact invariant subspace for $w>1$ in the spatial-temporal random RPM and 3PM (not at the DU limit).

\section{Mathematical proofs}
\subsection{Proof of $\abs{\sigma_{\text{spectrum}}(T_w)}\leq1$}
\subsubsection{Large $\tau$ behaviour of $T_w$}\label{Sec:Eigen}
In the main text, we show that the \LLG\ $T_w$ is generally not contracting, so one does not know  a priori that its eigenvalues lie in the complex unit disk. Here, we provide proof that the eigenvalues indeed lie within the complex unit disk. 

Let us first prove the following lemma: if the maximal modulus of the eigenvalue of a matrix $T$ is $z_>$, then for two arbitrary sets of complete basis $\bra{v_\alpha^{\text{L}}}$ and $\ket{v_\beta^{\text{R}}}$, then
\begin{align}
	\limsup_{\tau\to\infty}\frac{\abs{\mel{v_\alpha^{\text{L}}}{T^\tau}{v_\beta^{\text{R}}}}}{\tau^{\varphi(z_>)}z_>^\tau}<+\infty,
\end{align}
and there exist a pair of basis vectors that makes this limit nonzero. Here, $\varphi(z)$ is determined by
\begin{align}
	\varphi(z):=\max_{i:\abs{z_i}=\abs{z}}\{N_i-1\},
\end{align}
where $z_i$ and $N_i$ are respectively the eigenvalue and the dimension of the $i$th Jordan block of $T$.

Suppose the Jordan normal form of $T$ is given by
\begin{align}
	T=S^{-1}JS=S^{-1}\qty[\sum_i(D_{i}+M_{i} )]S,
\end{align}
where $D_i$ and $M_i$ are the diagonal and off-diagonal parts of the $i$th Jordan block, respectively. As the basis is arbitrary, the invertible matrix can be absorbed into the basis, and thus, we set $T=J$ without loss of generality.
Therefore, for $\tau>\varphi(z_>)$, we have
\begin{align}
	T^\tau=\sum_i\sum_{j=0}^{N_i-1}\binom{\tau}{j}(D_i)^{\tau-j}(M_i)^j \;.
\end{align}
Consequently, the limit becomes
\begin{widetext}
\begin{align}\label{Eq:Limit}
	\limsup_{\tau\to\infty}\frac{\abs{\mel{v_\alpha^{\text{L}}}{T^\tau}{v_\beta^{\text{R}}}}}{\tau^{\varphi(z_>)}z_>^\tau}
	=\limsup_{\tau\to\infty}\abs{\sum_i\sum_{j=0}^{N_i-1}\frac{z_>^{-j}}{\tau^{\varphi(z_>)}}\binom{\tau}{j}\qty[\frac{z_i}{z_>}]^{\tau-j}\mel{v_\alpha^{\text{L}} }{[M_{z_i}]^j}{v_\beta^{\text{R}} }}
	=\limsup_{\tau\to\infty}\frac{\abs{\mel{v_\alpha^{\text{L}} }{Q(\tau)}{v_\beta^{\text{R}} }}}{\varphi(z_>)!\,z_>^{\varphi(z_>)}},
\end{align}
\end{widetext}
where $Q(\tau)$ denotes
\begin{align}\label{Eq:Tlim}
	Q(\tau)=\sum_{\substack{i:\abs{z_i}=z_>,\\ ~N_i=\varphi(z_>)+1}}\qty(z_i/z_>)^{\tau-\varphi(z_>)}(M_i)^{\varphi(z_>)}.
\end{align}
Because $\norm{Q(\tau)}=1$ and all finite-dimensional norms are equivalent, we have that Eq.~\eqref{Eq:Limit} is always finite and
\begin{align}
	\limsup_{\tau\to\infty}\sum_{\alpha,\beta}\frac{\abs{\mel{v_\alpha^{\text{L}}}{T^\tau}{v_\beta^{\text{R}}}}}{\tau^{\varphi(z_>)}z_>^\tau}>0,
\end{align}
meaning that the limit is nonzero for at least one pair of $\alpha,\beta$. 

\subsubsection{Asympototic behavior of the singular value}
Note that ``$\limsup$'' can be replaced by ``$\lim$'' if there is only one Jordan block contributing in Eq.~\eqref{Eq:Tlim}, because
\begin{align}
	\lim_{\tau\to\infty}\frac{\varphi(z_>)! T^\tau}{\tau^{\varphi(z_>)}z_>^{\tau - \varphi(z_>)}}=(M_{z_2})^{\varphi(z_>)}=\ket{z_{>}^{\text{R}}}\bra{z_{>}^{\text{L}}},
\end{align}
where $\bra{z_{>}^{\text{L}}},\ket{z_{>}^{\text{R}}}$ are the left and right eigenstates of eigenvalue $z_>$. What is more, we also have
\begin{align}
	\varphi(z_{>})=a(z_>)-1,
\end{align}
where $a(z)$ is the algebraic multiplicity of $z$.
In this case, the singular value $\lambda_\tau$ is asymptotically given by
\begin{align}
	\lim_{\tau\to\infty}\frac{\varphi(z_>)! \lambda_\tau}{\tau^{\varphi(z_>)}z_>^{\tau-\varphi(\tau)}\norm{\ket{z_>^{\text{R}}}}\norm{ \ket{ z_>^{\text{L}}} } }=1,
\end{align}
and the singular state satisfies
\begin{align}\label{Eq:SingularState}
	\lim_{\tau\to\infty}\ket{\lambda_\tau^{\text{L}}}=\frac{\ket{z_{>}^{\text{R}}}}{\norm{z_{>}^{\text{R}}}},\quad
	\lim_{\tau\to\infty}\ket{\lambda_\tau^{\text{R}}}=\frac{\ket{z_{>}^{\text{L}}}}{\norm{z_{>}^{\text{L}}}}.
\end{align}
Applying the lemma to the \LLG\ $F_w$, we expect that the OTOC should scale as $\tau^{\varphi[z_2(w),w]}z_2(w)^\tau$ at large $\tau$, where $z_2(w)$ is the leading eigenvalue of $F_w$ and equivalently, the subleading eigenvalue of $T_w$.

\subsubsection{Proof of $\mel{v_{\mathrm{L}}}{T_w^\tau}{v_{\mathrm{R}}}<\infty$}\label{sec:upper_bound}
Now, let us come back to the \LLG.
If the leading eigenvalue of $T_w$ is greater than 1 and there are two sets of complete orthogonal basis, there exist two basis vectors satisfying
\begin{align}\label{Eq:Divergence}
	\limsup_{\tau\to\infty}\abs{\mel{v_{\mathrm{L}}}{T_w^\tau}{v_{\mathrm{R}}}}=\infty.
\end{align}
Thus, there exists $\tau_0>0$ such that $\abs{\mel{v_{\mathrm{L}}}{T_w^{\tau_0}}{v_{\mathrm{R}}}}>1$. 

On the other hand, by using $\sigma_\alpha$ ($\alpha=1,2,\cdots,q^2$), one can construct a complete orthogonal basis with $\bigotimes_{i=1}^w \qty[\sigma_{\alpha_i}\Otimesk\sigma_{\beta_i}]$.
Here, the operator norm of the basis is normalized, $\norm{\sigma_\alpha}=1$.
The desired pair of basis is thus given by
\begin{align}
	v_{\mathrm{L}}&=\frac1{q^{w/2}}\bigotimes_{i=1}^w\mathcal{T}_\text{k}\qty[\sigma_{a_i}\Otimesk\sigma_{c_i}],\nonumber\\
	v_{\mathrm{R}}&=\frac1{q^{w/2}}\bigotimes_{i=1}^w\qty[\sigma_{b_i}\Otimesk\sigma_{d_i}],\nonumber
\end{align}
where $\mathcal{T}_\text{k}$ represents the translation in the replicated space. 

Next, we will prove that $\abs{\mel{v_{\mathrm{L}}}{T_w^{\tau_0}}{v_{\mathrm{R}}}}$ must be equal or less than $1$, resulting in the contradiction. Essentially, we will construct four operators $A,B,C,D$ with $\norm{A},\norm{B},\norm{C},\norm{D}\leq1$, and an unitary evolution $U(t)$ that satisfying
\begin{align}\label{Eq:Construction}
	\expval{A(t)BC(t)D}=\mel{v_{\mathrm{L}}}{T_w^\tau}{v_{\mathrm{R}}},
\end{align}
where $A(t)=U(t)^\dag A U(t)$ and $C(t)=U(t)^\dag C U(t)$. If we find such construction, then we have
\begin{align}
	\mel{v_{\mathrm{L}}}{T_w^\tau}{v_{\mathrm{R}}}&=\expval{A(t)BC(t)D}\nonumber\\
	&\leq\norm{A(t)BC(t)D}\nonumber\\
	&\leq\norm{A(t)}\norm{B}\norm{C(t)}\norm{D}\leq 1.
\end{align}

\begin{figure}[h]
\center
\includegraphics[width=0.9\columnwidth]{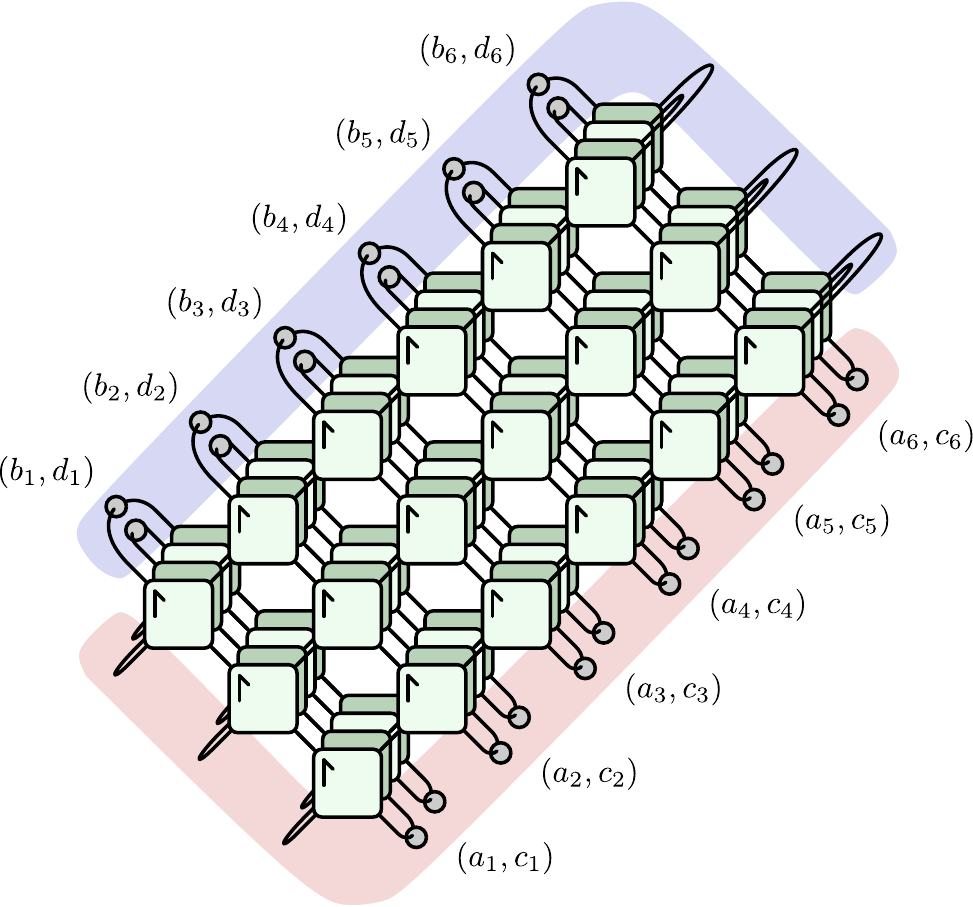}
\caption{\label{Fig:Construction}An illustration for $\mel{v_{\mathrm{L}}}{T_w^\tau}{v_{\mathrm{R}}}$ with operators $A$ and $C$ highlighed in red, and $B$ and $D$ highlighted in blue. }
\end{figure}

Indeed, we can find such construction. $A,C$ are given by
\begin{align}
	A=\mathbb{I}^{\tau}\otimes\bigotimes_{i=1}^{w}\sigma_{a_i},\quad 
	C=\mathbb{I}^{\tau}\otimes\bigotimes_{i=1}^{w}\sigma_{c_i},
\end{align}
and they are supported on sites $x\in[\tau,w+\tau]$. $B,D$ are given by
\begin{align}
	B=\bigotimes_{i=1}^{w}\sigma_{b_i}\otimes\mathbb{I}^{\tau},\quad 
	D=\bigotimes_{i=1}^{l}\sigma_{d_i}\otimes\mathbb{I}^{\tau},
\end{align}
and they are supported on sites $x\in[1,w]$. For the evolution operator, we first define the evolution of each time step $\tilde U(t)$
\begin{align}
	\tilde U(t)=\mathbb{I}^{\abs{t-\tau}}\otimes u^{w+\tau-\abs{t-\tau}-\abs{t-w}} \otimes \mathbb{I}^{\abs{t-w}}.
\end{align}
The evolution operator is then defined as $U(t)=\prod_{t=1}^{w+\tau}\tilde U(t)$ as illustrated by Fig.~\ref{Fig:Construction}.
It is easy to verify that $\norm{A},\norm{B},\norm{C},\norm{D}\leq1$, and that Eq.~\eqref{Eq:Construction} is satisfied. This completes the proof.

\subsubsection{Multiplicity of leading eigenvalues}

From the construction, we also know that
\begin{align}
	\limsup_{\tau\to\infty}\abs{\mel{v_{\mathrm{L}}}{T_w^\tau}{v_{\mathrm{R}}}} < \infty 
\end{align}
is always finite. Hence, the modulus of the leading eigenvalue is less than or equal to $1$. Moreover, the algebraic multiplicity of the eigenvalue with modulus $1$ must equal the geometric multiplicity, or the OTOC grows algebraically.

\subsection{Proof of $\norm{T_w}\leq q$}\label{Sec:Norm}
Following from the construction in Section~\ref{sec:upper_bound}, we know that for an arbitrary vector $\ket{v_{\mathrm{R}}}$, $T_w\ket{v_{\mathrm{R}}}$ can be expressed as
\begin{align}
	\norm{T_w\ket{v_{\mathrm{R}}}}=\norm{P_r[\tilde U(\mathbb{I}\otimes v_{\mathrm{R}}) \tilde U^\dag]}_\text{F},
\end{align}
where $\norm{\cdot}_\text{F}$ is the Frobenius norm, and $\tilde U=U\Otimesk U$ with $U$ given by the construction. Here, we consider $v_{\mathrm{R}}$ as an operator acting on a Hilbert space of dimension $q^{2w}$, and $\ket{0},\ket{1}$ correspond to the identity $\mathbb{I}$ and the swap $\mathbb{S}$, respectively.
Additionally, $P_r$ is a projection operator that leaves only the component with the swap operator on the rightmost site, i.e. $\hat O\otimes \mathbb{S}$. Therefore,
we have
\begin{align}\label{Eq:UpperBound}
	\norm{T_w\ket{v_{\mathrm{R}}}}&\leq \norm{\tilde U(\mathbb{I}\otimes v_{\mathrm{R}}) \tilde U^\dag}_\text{F}\nonumber\\
	&=\norm{\mathbb{I}\otimes v_{\mathrm{R}}}_\text{F}=\norm{\mathbb{I}}_\text{F}\norm{v_{\mathrm{R}}}_\text{F}=q\norm{\ket{v_{\mathrm{R}}}},
\end{align}
which completes the proof. This bound can be further improved to $\norm{T_w}\leq1$ for the dual-unitary case \cite{bertini_op_2020}.
This means $\alpha\leq1$, and the lower bound of $\alpha$ is trivial because the leading eigenvalue of $T_w$ is $1$.

On the other hand, the reducibility (see Fig.~1 in the main text) implies that a longer \LLG\ inherits the spectrum of a shorter one. Furthermore, we can also conclude that $\norm{T_w}$ increases monotonically with respect to $w$. Therefore, $\alpha=\lim_{w\to\infty}\ln(\norm{T_w})/\ln q$ exists.

\subsection{Proof of  $z_2(w)=z_2(w=1)$}\label{Sec:z2_proof}

For the purpose of the proof, we introduce the generalized Pauli matrices, defined as
\begin{align}
	\sigma^{j,k}=\sum_{m=0}^{q-1}\omega^{jm}\dyad{m+k}{m},
\end{align}
where $j,k=0,\cdots,q-1$, and $\omega=e^{2\pi i/q}$. For convenience, we also use $\mu=j+qk$ to label $\sigma^{j,k}$. The generalized Pauli matrices satisfy $\trace[(\sigma^\mu)^\dag\sigma^\nu]=q\delta_{\mu\nu}$, and $\sigma^{j,k}\sigma^{j^\prime,k'}=\omega^{jk'}\sigma^{(j+j'),(k+k')}$. A useful identity is that for $\nu\neq 0$,
\begin{align}\label{Eq:Pauli}
	\sum_{\mu=1}^{q^2-1}\norm{[\sigma^\mu,\sigma^\nu]}_\text{F}^2=2q^3.
\end{align}
What is more, we generalized OTOC for the generalized Pauli matrices, which are unitary but non-Hermitian,
\begin{align}
	\tilde C_{\mu\nu}(x,t)&=1-\expval{\sigma^{\mu}(0,t)^\dag\sigma^{\nu}(x,0)^\dag\sigma^{\mu}(0,t)\sigma^{\nu}(x,0)}.
\end{align}
Hence, we have $C_{\mu\nu}(w,\tau)=\mel{L_{\mu\nu}}{F_w^\tau}{R_{\mu\nu}}$, where
\begin{align}
	\bra{L_{\mu\nu}}&=\bra{1^{w-1}}\Otimes{}{r}\bra{1_{(\sigma^\nu)^\dag,\sigma^\nu}},\nonumber\\
	\ket{R_{\mu\nu}}&=\ket{0_{(\sigma^\mu)^\dag,\sigma^\mu}}\Otimes{}{r}\ket{0^{w-1}},
\end{align}
and
\begin{align}
	\mathrm{Re}[\tilde C_{\mu\nu}(x,t)]
	=\frac12\expval{[\sigma^{\mu}(0,t),\sigma^{\nu}(x,0)]^\dag[\sigma^{\mu}(0,t),\sigma^{\nu}(x,0)]}.
\end{align}

\subsubsection{Main proof}
\begin{theorem}
If there exist $\mu$ and $\nu$ that
\begin{align}\label{Eq:Condition}
	\liminf_{\tau\to\infty}\frac{\mathrm{Re}[C_{\mu\nu}(w,\tau)] }{\tau^{\varphi(w)} \abs{z_2(w)}^\tau}>0,
\end{align}
then $\abs{z_2(w)}=\abs{z_2(w-1)}$. 
\end{theorem}

First, we have
\begin{align}
	\liminf_{\tau\to\infty}\frac{\abs{C_{\mu\nu}(w,\tau)} }{\tau^{\varphi(w)} \abs{z_2(w)}^\tau}>0,\nonumber\\
	\limsup_{\tau\to\infty}\frac{\abs{C_{\mu\nu}(w,\tau)} }{\tau^{\varphi(w)} \abs{z_2(w)}^\tau}<\infty.
\end{align}

Let $P_2$ be the projector operator of the 2-point correlation function part of the OTOC \LLG, given by
\begin{align}
	P_2=\dyad{\mathbbm{I}^w}\Otimes{}{k}\overline{\dyad{\mathbb{I}^w}}+\overline{\dyad{\mathbb{I}^w}}\Otimes{}{k}\dyad{\mathbb{I}^w}
\end{align}
where the overline denotes the orthogonal complement of a projection operator. We have
\begin{align}
	P_2\ket{L}=P_2\ket{R}=0,\quad F_wP_2=P_2F_wP_2.
\end{align}

We define $\ket{R(\tau)}=\bar P_2F_w^\tau\ket{R}$, and have
\begin{align}
	X&=\limsup_{\tau\to\infty}\frac{\norm{\ket{L}} \norm{\ket{R(\tau)}} }{\abs{\braket{L}{R(\tau)}}}\nonumber\\
	&\leq
	\limsup_{\tau\to\infty}\frac{\norm{F_w^\tau}\norm{\ket{L}} \norm{\ket{R}} }{\abs{\braket{L}{R(\tau)}}}<\infty.
\end{align}
Further, we have
\begin{align}
	\braket{L}{R(\tau+N)}&=\mel{L}{P_2F_w^{\tau+N}}{R}\nonumber\\
	&=\mel{L}{F_w^{N}(P_2+\bar P_2)F_w^\tau}{R}\nonumber\\
	&=\mel{L}{P_2F_w^{N}P_2F_w^\tau}{R}+\mel{L}{F_w^{N}\bar P_2F_w^\tau}{R}\nonumber\\
	&=\mel{L}{F_w^{N}\bar P_2F_w^\tau}{R}=\mel{L}{F_w^{N}}{R(\tau)}.
\end{align}
We assume $\abs{z_2(w)}>\abs{z_2(w-1)}$ henceforth. By defining the projection operator $P_w=[\ket{0^{w-1}}\bra{0^{w-1}}/q^{w-1}]\otimes\mathbb{I}$, we can obtain
\begin{align}\label{Eq:Transport}
	\limsup_{\tau\to\infty}\frac{\norm{\ket{R(\tau)}-\ket{R'(\tau)}} }{\norm{\ket{R(\tau)}}}=0,
\end{align}
where $\ket{R'(\tau)}=P_w\ket{R(\tau)}=\ket{0^{w-1}}\otimes\ket*{\tilde R(\tau)}$. The proof of this equation is left to the next section. Hence, we also have
\begin{align}
	\limsup_{\tau\to\infty}\frac{\norm{\ket{R(\tau)}-\ket{R'(\tau)}} }{\abs{\braket{L}{R(\tau)}}}=0,
\end{align}
Further, there is an integer $N$ that $\norm{F_{w=1}^N}<\abs{z(w)}^N/(2X)$, because $\lim_{\tau\to\infty}\norm{F_{w=1}^\tau}^{1/\tau}=\abs{z_2(1)}<\abs{z_2(w)}$.

For $\tau>N$, we have
\begin{align}
	\frac{\abs{C_{\mu\nu}(w,\tau+N)}}{\abs{C_{\mu\nu}(w,\tau)}}\leq\, &
	\frac{\norm{\ket{L}}\norm{F_w^N}\norm{\ket{R(\tau)}-\ket{R'(\tau)}} }{\abs{\braket{L}{R(\tau)}}}\nonumber\\
	&+\frac{\abs{\mel{L}{F_w^N}{R'(\tau)}} }{\abs{\braket{L}{R(\tau)}}}\nonumber\\
	\leq\, &
	\frac{2^Nq^{wN}\norm{\ket{L}}\norm{\ket{R(\tau)}-\ket{R'(\tau)}} }{\abs{\braket{L}{R(\tau)}}}\nonumber\\
	&+\frac{\abs{\mel{\tilde L}{F_1^N}{\tilde R(\tau)}} }{\abs{\braket{L}{R(\tau)}}}\nonumber\\
	\leq\, &
	2^Nq^{wN}\frac{\norm{\ket{L}}\norm{\ket{R(\tau)}-\ket{R'(\tau)}} }{\abs{\braket{L}{R(\tau)}}}\nonumber\\
	&+\frac{\abs{z_2(w)}^N}{4X}\frac{\norm*{\ket*{\tilde L}}\norm*{\ket*{\tilde R(\tau)}}  }{\abs{\braket{L}{R(\tau)}}}\nonumber\\
	\leq\, &
	2^Nq^{wN}\frac{\norm{\ket{L}}\norm{\ket{R(\tau)}-\ket{R'(\tau)}} }{\abs{\braket{L}{R(\tau)}}}\nonumber\\
	&+\frac{\abs{z_2(w)}^N}{2X}\frac{\norm{\ket{L}}\norm{\ket{R(\tau)}}  }{\abs{\braket{L}{R(\tau)}}}.
\end{align}
Here, we used $\norm{F_w}\leq q^w+\norm{T_w}\leq 2q^w$ for the second inequality.
Therefore, we have
\begin{align}
	\abs{z_2(w)}^N\leq\limsup_{\tau\to\infty}\frac{\abs{C_{\mu\nu}(w,\tau+N)}}{\abs{C_{\mu\nu}(w,\tau)}}\leq \frac{\abs{ z_2(w)}^N}{2}, 
\end{align}
resulting in a contradiction.

Hence, if the premise is always satisfied, we obtain the following corollary,
\begin{corollary}
If there exist $\mu_w$ and $\nu_w$ for all $w$ that
\begin{align}
	\liminf_{\tau\to\infty}\frac{\mathrm{Re}[C_{\mu_w\nu_w}(w,\tau)]}{\tau^{\varphi(w)} \abs{z_2(w)}^\tau}>0,
\end{align}
then $z_2(w)=z_2(w=1)$. 
\end{corollary}

\subsubsection{Proof of Equation~(\ref{Eq:Transport}) }
From Eq.~\eqref{Eq:Condition} and assuming $\abs{z_2(w)}>\abs{z_2(w-1)}$, we know that
\begin{align}\label{Eq:Premise}
	\lim_{\tau\to\infty}\frac{\sum_{w'=1}^{w-1}\sum_{\alpha} \mathrm{Re}[C_{\mu\alpha}(w',\tau)]}{\sum_{\alpha}\mathrm{Re}[C_{\mu\alpha}(w,\tau)]}=0.
\end{align}
The evolution of the operator can be explicitly written as
\begin{align}
	\sigma^\mu(0,t)=\,&\sideset{}{'}\sum_{\alpha_1,\cdots,\alpha_{w-1}}\sum_{\alpha_w}\sigma_{\alpha_1}\otimes\cdots\otimes\sigma_{\alpha_w}\otimes A_{\alpha_1\cdots\alpha_w}\nonumber\\
	&+\sideset{}{'}\sum_{\alpha_{w}}\mathbb{I}^{w-1}\otimes\sigma_{\alpha_w}\otimes A_{\alpha_w}\nonumber\\
	&+\mathbb{I}^{w}\otimes A,
\end{align}
where $\sideset{}{'}\sum$ denotes that at least one index is nonzero. 
Using singular value decomposition, this can be reexpressed as
\begin{align}
	\sigma^\mu(0,t)=\,&\sum_{\gamma}S_\gamma\otimes A'_{\gamma}+\sum_{\alpha=1}^{q^2-1}\mathbb{I}^{w-1}\otimes S_\alpha\otimes A_{\alpha}'+\mathbb{I}^{w}\otimes A,
\end{align}
where $S_\gamma,S_\alpha,A'_\gamma,A'_\alpha$ are orthogonal, and we adopt the following normalization $\norm{S_\alpha}_\text{F}^2=q$ and $\norm{S_\gamma}_\text{F}^2=q^{w}$.
Hence, we have
\begin{align}
	q^{\tau}&=\sideset{}{'}\sum_{\alpha_1,\cdots,\alpha_{w-1}}\sum_{\alpha_w}\norm{A_{\alpha_1\cdots\alpha_w}}_\text{F}^2+\sideset{}{'}\sum_{\alpha_w}\norm{A_{\alpha_w}}_\text{F}^2+\norm{A}_\text{F}^2\nonumber\\
	&=\sum_{\gamma}\norm{A_{\gamma}'}_\text{F}^2+\sum_{\alpha=1}^{q^2-1}\norm{A_{\alpha}'}_\text{F}^2+\norm{A}_\text{F}^2.
\end{align}
Hence, we obtain the expansion for the OTOC with $w'<w$,
\begin{align}
	\sum_{\alpha}\mathrm{Re}[C_{\mu\alpha}(w',\tau)]=\frac{q^2}{q^{\tau}}\sum_{\substack{\alpha_1,\cdots,\alpha_w\\ \alpha_{w'}\neq0}}\norm{A_{\alpha_1\cdots\alpha_w}}_\text{F}^2,
\end{align}
where we used Eq.~\eqref{Eq:Pauli},
and
\begin{align}
	\sum_{\alpha}\mathrm{Re}[C_{\mu\alpha}(w,\tau)]=\,&\frac{q^2}{q^{\tau}}\sideset{}{'}\sum_{\alpha_1,\cdots,\alpha_{w-1}}\sideset{}{'}\sum_{\alpha_{w}}\norm{A_{\alpha_1\cdots\alpha_w}}_\text{F}^2\nonumber\\
	&+\frac{q^2}{q^{\tau}}\sideset{}{'}\sum_{\alpha_{w}}\norm{A_{\alpha_w}}_\text{F}^2.
\end{align}
Following this, we obtain
\begin{align}
	\sum_{w'=1}^{w-1}\sum_{\alpha}\mathrm{Re}[C_{\mu\alpha}(w',\tau)]\geq \frac{q^2}{q^{\tau}}\sideset{}{'}\sum_{\alpha_1,\cdots,\alpha_{w-1}}\sum_{\alpha_w}\norm{A_{\alpha_1\cdots\alpha_w}}_\text{F}^2,
\end{align}
and incorporating Eq.~\eqref{Eq:Premise}, we deduce
\begin{align}
	\lim_{\tau\to\infty}\frac{\sum_{\gamma}\norm{A_{\gamma}'}_\text{F}^2}
	{\sum_{\alpha=1}^{q^2-1}\norm{A_{\alpha}'}_\text{F}^2}=0.
\end{align}
Further, we know
\begin{align}
	q^{w/2}T_w^\tau\ket{R}=\,&\frac{1}{q^{\tau}}\trace_{\tau}[\sigma^\mu(0,t)^\dag\Otimes{}{k}\sigma^\mu(0,t)]\nonumber\\
	=\,&\frac{\norm{A}_\text{F}^2}{q^{\tau}}\,\mathbb{I}^{w}\Otimes{}{k}\mathbb{I}^{w}\nonumber\\
	&+\mathbb{I}^{w}\Otimes{}{k}\mathbb{I}^{w-1}\otimes \sum_{\alpha=1}^{q^2-1} \frac{\trace[A^\dag A'_\alpha]}{q^{\tau}}\,S_\alpha\nonumber\\
	&+\mathbb{I}^{w-1}\otimes \sum_{\alpha=1}^{q^2-1} \frac{\trace[(A'_\alpha)^\dag A]}{q^{\tau}}\,S_\alpha^\dag\Otimes{}{k}\mathbb{I}^{w}\nonumber\\
	&+\mathbb{I}^{w}\Otimes{}{k}\sum_\gamma \frac{\trace[A^\dag A'_\gamma]}{q^{\tau}}\,S_\gamma
	+\sum_\gamma \frac{\trace[(A'_\gamma)^\dag A]}{q^{\tau}}\,S_\gamma^\dag\Otimes{}{k}\mathbb{I}^{w}
	\nonumber\\
	&+\mathbb{I}^{w-1}\otimes\sum_{\alpha=1}^{q^2-1}\frac{\trace[(A'_\alpha)^\dag A'_\gamma]}{q^{\tau}}\, S_\alpha^\dag\Otimes{}{k}\sum_\gamma S_\gamma\nonumber\\
	&+\sum_\gamma S_\gamma^\dag\Otimes{}{k}\mathbb{I}^{w-1}\otimes\sum_{\alpha=1}^{q^2-1}\frac{\trace[(A'_\gamma)^\dag A'_\alpha]}{2^{\tau}}\, S_\alpha\nonumber\\
	&+\sum_{\alpha=1}^{q^2-1}\frac{\norm{A'_\alpha}_\text{F}^2}{q^{\tau}}\,  S_\alpha^\dag\otimes S_\alpha+\sum_\gamma \frac{\norm{A'_\gamma}_\text{F}^2}{q^{\tau}}\, S_\gamma^\dag\otimes S_\gamma.
\end{align}
Hence,
\begin{align}
	&q^{w/2}\ket{R(\tau)}\nonumber\\
	=\,&q^{w/2}\bar P_2T_w^\tau\ket{R}-q^{w/2}\ket{0^w}\nonumber\\
	=\,&-\qty(\sum_{\gamma}\norm{A_{\gamma}'}_\text{F}^2+\sum_{\alpha=1}^{q^2-1}\norm{A_{\alpha}'}_\text{F}^2)\,\mathbb{I}^{w}\Otimes{}{k}\mathbb{I}^{w}\nonumber\\
	&+\mathbb{I}^{w-1}\otimes\sum_{\alpha=1}^{q^2-1}\frac{\trace[(A'_\alpha)^\dag A'_\gamma]}{q^{\tau}}\, S_\alpha^\dag\Otimes{}{k}\sum_\gamma S_\gamma\nonumber\\
	&+\sum_\gamma S_\gamma^\dag\Otimes{}{k}\mathbb{I}^{w-1}\otimes\sum_{\alpha=1}^{q^2-1}\frac{\trace[(A'_\gamma)^\dag A'_\alpha]}{2^{\tau}}\, S_\alpha\nonumber\\
	&+q^{w/2}\ket{0^w}\otimes\sum_{\alpha=1}^{q^2-1}\frac{\norm{A'_\alpha}_\text{F}^2}{q^{\tau}}\,  S_\alpha^\dag\Otimes{}{k} S_\alpha+\sum_\gamma \frac{\norm{A'_\gamma}_\text{F}^2}{q^{\tau}}\, S_\gamma^\dag\Otimes{}{k} S_\gamma.
\end{align}
We do the following decomposition,
\begin{align}
	\ket{R(\tau)}=\ket{R_*(\tau)}+\ket{R'(\tau)}.
\end{align}
The norm of the two terms can be estimated by
\begin{align}
	\frac{\braket{R'(\tau)}}{q^{w-2\tau}}\geq\sum_{\alpha=1}^{q^2-1}\norm{A'_\alpha}_\text{F}^4\geq\frac1{q^2-1}\qty[ \sum_{\alpha=1}^{q^2-1}\norm{A'_\alpha}_\text{F}^2 ]^2
\end{align}
and
\begin{align}
	&\frac{\braket{R_*(\tau)}}{q^{w-2\tau}}\nonumber\\
	=\,&2\sum_{\alpha=1}^{q^2-1}\sum_{\gamma}\abs{\trace[(A'_\alpha)^\dag A'_\gamma]}^2+\sum_{\gamma}\norm{A'_\gamma}_\text{F}^4\nonumber\\
	\leq\,& 2\sum_{\alpha=1}^{q^2-1}\sum_{\gamma}\norm{A'_\alpha}_\text{F}^2\norm{A'_\gamma}_\text{F}^2+\qty[\sum_{\gamma}\norm{A'_\gamma}_\text{F}^2]^2\nonumber\\
	=\,&2\qty[\sum_{\alpha=1}^{q^2-1}\norm{A'_\alpha}_\text{F}^2]\qty[\sum_{\gamma}\norm{A'_\gamma}_\text{F}^2]+\qty[\sum_{\gamma}\norm{A'_\gamma}_\text{F}^2]^2.
\end{align}
This implies that
\begin{align}
	\lim_{\tau\to\infty}\frac{\norm{\ket{R_*(\tau)}}}{\norm{\ket{R'(\tau)}}}=0,
\end{align}
and thus,
\begin{align}
	\lim_{\tau\to\infty}\frac{\norm{\ket{R(\tau)}-\ket{R'(\tau)}}}{\norm{\ket{R(\tau)}}}=0,
\end{align}

\subsection{Spatial-temporal random Haar-random model}\label{Sec:Case A HRM}

In this section, we will analytically solve the eigenvalues of the ensemble-averaged \LLG\ for the spatial-temporal random HRM. Essentially, we adopt a particular similar transformation that makes the \LLG\ an upper-triangular matrix, and consequently, the eigenvalues are given by the diagonal entries.

\subsubsection{Similar transformation}

Note that the basis is not orthogonal here, and it satisfies
\begin{align}
	\braket{i}{j}=q\delta_{ij}+\delta_{i\bar j},
\end{align}
where $\bar i=1-i$. To simplify the proof, we convert Eq.~\eqref{Eq:Original Tensor} into the biorthogonal basis,
\begin{align}
	\overline{\UU_2}=M_{j_1 j_2}^{i_1i_2}\ket{i_1}\rket{j_1'}\rbra{i_2}\bra{j_2'},
\end{align}
where $\ket{i'}$ is defined by $\bra{i}\ket{j'}=\delta_{ij}$. Hence, we have $\ket{i}=U_i^j\ket{j'}$ with $U_i^j=q\delta_i^j+\delta_{\bar i}^{j}$, and
\begin{align}
	M_{j_1 j_2}^{i_1i_2}=U_{j_1}^{k}U_{j_2}^{k}\delta^{i_1i_2}q^2\text{Wg}(i_1^{-1}k,q^2).
\end{align}
If $j_1=j_2=i_1$, we have
\begin{align}\label{Eq:Tensor}
	M_{00}^{00}=q^2q^2\frac1{q^4-1}-q^2\frac1{q^2(q^4-1)}=1.
\end{align}
If $j_1\neq j_2$, we have
\begin{align}
	M_{01}^{00}=q^3\frac1{q^4-1}-q^3\frac1{q^2(q^4-1)}=\frac{q}{q^2+1}.
\end{align}
To summarise, the nonvanishing entries of $\tilde M$ are
\begin{align}\label{Eq:Tensor}
	&M_{01}^{00}=M_{10}^{00}=M_{01}^{11}=M_{01}^{11}=\frac q{q^2+1},\\
	&M_{00}^{00}=M_{11}^{11}=1.
\end{align}
Obviously, $M$ satisfies the following symmetry
\begin{align}\label{Eq:Symmetry}
	M_{ij}^{kl}=M_{\bar i\bar j}^{\bar k \bar l}.
\end{align}

The light-like (LL) generator is given by
\begin{align}\label{Eq:LL}
	T_{(j_1j_2\cdots j_w)}^{(i_1i_2\cdots i_w)}
	=&M_{0j_1}^{i_1\mu_1}M_{\mu_1j_2}^{i_2\mu_2}\cdots M_{\mu_{w-1}j_w}^{i_w\alpha}(\alpha|1)\nonumber\\
	=&M_{0j_1}^{i_1\mu_1}M_{\mu_1j_2}^{i_2\mu_2}\cdots M_{\mu_{w-1}j_w}^{i_w0}\nonumber\\
	&+q M_{0j_1}^{i_1\mu_1} M_{\mu_1j_2}^{i_2\mu_2}\cdots M_{\mu_{w-1}j_w}^{i_w1}.
\end{align}
Thus, the \LLG\ contains two terms for $\alpha=0$ and $\alpha=1$. Moreover, because of Eq.~\eqref{Eq:Tensor}, the tensor contraction can be reduced to a simple product,
\begin{align}
	M_{0j_1}^{i_1\mu_1}M_{\mu_1j_2}^{i_2\mu_2}\cdots M_{\mu_{w-1}j_w}^{i_w\alpha}
	=M_{0j_1}^{i_1i_1}\times M_{i_1j_2}^{i_2i_2}\times\cdots\times M_{i_{w-1}j_w}^{i_w\alpha},
\end{align}
where the right-hand side has no Einstein summation.

The next step is to reorder the basis so that the \LLG\ becomes upper triangular. Before the proof, it is useful to introduce three propositions of the \LLG.
\begin{proposition}\label{Prop:1}
$T_{(0j_2\cdots j_w)}^{(1i_2\cdots i_w)}=0$.
\end{proposition}
This is because $M_{00}^{11}=0$ according to Eq.~\eqref{Eq:Tensor}.

\begin{proposition}\label{Prop:2}
$T_{(0j_2\cdots j_w)}^{(0i_2\cdots i_w)}=T_{(j_2\cdots j_w)}^{(i_2\cdots i_w)}$.
\end{proposition}
This is because $M_{00}^{00}=1$ according to Eq.~\eqref{Eq:Tensor}.

\begin{proposition}\label{Prop:3}
If $T_{(0j_2\cdots j_w)}^{(0i_2\cdots i_w)}=0$, then $T_{(1\bar j_2\cdots \bar j_w)}^{(1\bar i_2\cdots \bar i_w)}=0$.
\end{proposition}
First, we have $T_{(0j_2\cdots j_w)}^{(0i_2\cdots i_w)}=T_{(j_2\cdots j_w)}^{(i_2\cdots i_w)}=0$.
As two terms in Eq.~\eqref{Eq:LL} are nonnegative, $T_{(j_2\cdots j_w)}^{(i_2\cdots i_w)}=0$ implies that
\begin{align}
	&M_{0j_2}^{i_2i_2}\times\cdots\times M_{i_{w-1}j_w}^{i_w0}\nonumber\\
	=& M_{0j_1}^{i_1i_1}\times M_{0j_2}^{i_2i_2}\times\cdots\times M_{i_{w-1}j_w}^{i_w1}=0.
\end{align}
Further, because of the symmetry given in Eq.~\eqref{Eq:Symmetry}, we obtain
\begin{align}
	&M_{01}^{11}\times M_{1\bar j_2}^{\bar i_2\bar i_2}\times\cdots\times M_{\bar i_{w-1}\bar j_w}^{\bar i_w1}\nonumber\\
	=& M_{01}^{11}\times M_{1\bar j_1}^{i_1i_1}\times M_{1\bar j_2}^{\bar i_2\bar i_2}\times\cdots\times M_{\bar i_{w-1}\bar j_w}^{\bar i_w0}=0.
\end{align}
Therefore, we have $T_{(1\bar j_2\cdots \bar j_L)}^{(1\bar i_2\cdots \bar i_L)}=0$.

Using these three propositions, we can construct a map $\sigma_w(m)=(i_1i_2\cdots i_w)$ where $m=0,\cdots,2^w-1$, satisfying
$T_{\sigma_w(n)}^{\sigma_w(m)}=0$ if $m>n$. We will do this by induction. For $w=1$, let
\begin{align}
	\sigma_1(0)=0,\quad\sigma_1(1)=0,
\end{align}
and then $T_{(0)}^{(1)}=0$ because of Prop.~\ref{Prop:1}. Supposing that we have such a map for $w-1$, then for $w$, we define
\begin{align}
	&\sigma_w(m)
	=\left(0~\sigma_{w-1}(m)\right),\nonumber\\
	&\sigma_w(2^{w-1}+m)=\left(1~\overline{\sigma_{w-1}(m)}\right),
\end{align}
for $m=0,1,\cdots,2^{w-1}-1$.

If $2^{w-1}>m>n$, then $T_{\sigma_w(n)}^{\sigma_w(m)}=T_{(0~\sigma_{w-1}(n))}^{(0~\sigma_{w-1}(m))}=T_{\sigma_{w-1}(n)}^{\sigma_{w-1}(m)}=0$ because of Prop.~\ref{Prop:2}.

If $m+2^{w-1}\geq2^{w-1}>n$, then $T_{\sigma_w(n)}^{\sigma_w(m+2^{w-1})}=T_{(0~\sigma_{w-1}(n))}^{(1~\overline{\sigma_{w-1}(m)})}=0$ because of Prop.~\ref{Prop:1}.

If $m+2^{w-1}>n+2^{w-1}\geq2^{w-1}$, then $T_{\sigma_w(n+2^{w-1})}^{\sigma_w(m+2^{w-1})}=T_{(1~\overline{\sigma_{w-1}(n)})}^{(1~\overline{\sigma_{w-1}(m)})}=0$ because of Prop.~\ref{Prop:3}.

Hence, the construction of $\sigma_w(m)$ is legitimate.

\subsubsection{Eigenvalues}
We have proved that $T_{(j_1j_2\cdots j_w)}^{(i_1i_2\cdots i_w)}$ is upper-triangular before reordering, and thus, the eigenvalues are given by $T_{(i_1i_2\cdots j_w)}^{(i_1i_2\cdots i_w)}$ for all $(i_1i_2\cdots i_w)$. That is, we need to evaluate
\begin{align}
	T_{(i_1i_2\cdots j_w)}^{(i_1i_2\cdots i_w)}
	= &M_{0i_1}^{i_1i_1} M_{i_1i_2}^{i_2i_2}\cdots M_{i_{w-1}0}^{00}\delta_{i_w,0}\nonumber\\
	&+q M_{0i_1}^{i_1i_1} M_{i_1i_2}^{i_2i_2}\cdots M_{i_{w-1}1}^{11}\delta_{i_w,1}.
\end{align}
Each $ M_{i_{k-1}i_k}^{i_ki_k}$ contributes $\frac{q}{q^2+1}$ if $i_{k-1}\neq i_k$, or $1$ if $i_{k-1}=i_k$. What is more, $i_w$ contributes $1$ for $i_w=0$, and $q$ for $i_w=1$. Therefore, it amounts to counting the number of domain walls in $0i_1i_2\cdots i_{w-1}0$ (PBC) and $0i_1i_2\cdots i_{w-1}1$ (twisted PBC). Here, we regard the length of these two chains as $w$.

It is readily to prove that PBC only has even numbers of domain walls, and that twisted PBC only has odd numbers. Further, we have the following recursive relation
\begin{align}
	f_{\text{PBC}}(w,n)&=f_{\text{PBC}}(w-1,n)+f_{\text{tPBC}}(w-1,n-1),\nonumber\\
	f_{\text{tPBC}}(w,n)&=f_{\text{tPBC}}(w-1,n)+f_{\text{PBC}}(w-1,n-1),
\end{align}
where $f_{\text{PBC/tPBC}}(w,n)$ is the number of states with $n$ domain walls in a chain of length $w$.
Using induction, one can prove that $f_{\text{PBC/tPBC}}(w,n)={w\choose n}$ for even or odd $n$ respectively.

In conclusion, the eigenvalues of the \LLG\ are
\begin{align}
\varepsilon_n=
\begin{cases}
	\left(\frac{q}{q^2+1}\right)^n & n \text{\ even},\\
	q\left(\frac{q}{q^2+1}\right)^n& n \text{\ odd},
\end{cases}
\end{align}
with algebraic multiplicity $w\choose n$. Here, $n$ takes value from $0,1,\cdots,w-1$.

\subsubsection{Geometric multiplicity of the subleading eigenvalue}
From the previous section, the subleading eigenvalue is $q^2/(q^2+1)$ with algebraic multiplicity $w$. In this section, we will prove that its geometric multiplicity is one. Considering the subspace $V_w$ spanned by all $\ket{0^m}\otimes\ket{1^{w-m}}$ with $m=0,1,\cdots,w$ is an invariant subspace of both $T_w$ and $T_w^\dag$. Within this subspace, we particularly choose the following orthonormal basis, $\ket{e_w}=\ket{0^w}/q^{w/2}$ and for $m=0,1,\cdots, w-1$. It is easy to prove that $T_w\ket{1^w}\in V_w$ using induction. To see it, supposing $T_w\ket{1^w}\in V_w$, we have for $\ket{1^{w+1}}$,
\begin{align}
	&\frac{q^2+1}q T_{w+1}\ket{1^{w+1}}\nonumber\\
	&=\ket{0}\otimes T_w\ket{1^w}+\ket{1}\otimes\rbra{1}\Bigodot{w}{r}\overline{\UU_2}\rket{1}\ket{1^{w}}\nonumber\\
	&=\ket{0}\otimes T_w\ket{1^w}+\ket{1^{w+1}}.
\end{align}
As $\ket{0}\otimes T_w\ket{1^w}$ and $\ket{1^{w+1}}$ belong to $V_{w+1}$, we have $T_{w+1}\ket{1^{w+1}}\in V_{w+1}$. Further, $T_w\ket{1^w}\in V_w$ implies $T_{w+m}\ket{0^m}\otimes\ket{1^w}\in V_{w+m}$ because of the reducibility, and thus,  $V_w$ is an invariant subspace of both $T_w$ and $T_w^\dag$. Within this subspace, we choose the following orthonormal basis, $\ket{e_w}=\ket{0^w}/q^{w/2}$ and for $m=0,1,\cdots, w-1$
\begin{align}
	\ket{e_m}=\ket{0^{m}}\otimes\frac{\ket{v}}{q^{w/2}\sqrt{q^2-1}}\otimes\ket{1^{w-m-1}},
\end{align}
where $\ket{v}=q\ket{1}-\ket{0}$. Note that in this basis, the $\ket{1}$ is given by
 \begin{align}
 	q^{w/2}\ket{1^w}=[\chi(w),\chi(w-1),\cdots,\chi(1),1]^\top,
 \end{align}
 where $\chi(j)=q^{j-1}\sqrt{q^2-1}$. 
Because $\braket{0}{v}=0$, we have $\mel{e_m}{T_w}{e_n}=0$ if {m<n}, meaning that $T_w$ is lower-triangular in this basis. By direct calculation, we also have $\mel{e_w}{T_w}{e_w}=1$, $\mel{e_m}{T_w}{e_m}=q^2/(q^2+1)$ for $m\leq w-1$, and $\mel{e_{m+1}}{T_w}{e_{m}}=q^3/(q^2+1)^2$ for $m\leq w-2$. Hence, the subspace contains all the Jordan blocks (if there are more than one) of the subleading eigenvalue. What is more, as the $\mel{e_{m+1}}{T_w}{e_m}\neq 0$ for $m\leq w-2$ means that the geometric multiplicity of the subleading eigenvalue is one. This is because if there is more than one eigenvector, then there must be an eigenvector $\ket{\psi}$ that $\braket{e_m}{\psi}$ for some $m\leq w-2$. Suppose that $m_0\leq w-2$ is the smallest one among such $m$-s, and that $\braket{m_0}{\psi}=1$. We then have
\begin{align}
	&\frac{q^2}{q^2+1}\braket{e_{m_0+1}}{\psi}=\mel{e_{m_0+1}}{T_w}{\psi}\nonumber\\
	&=\mel{e_{m_0+1}}{T_w}{e_{m_0}}+\frac{q^2}{q^2+1}\braket{e_{m_0+1}}{\psi},
\end{align}
leading to $\mel{e_{m_0+1}}{T_w}{e_{m_0}}=0$. Therefore, we prove that the temporal-spatial random HRM follows the two conjectures in the main text.

\subsubsection{Leading singular value around the butterfly cone}

Because $\ket{1^w},\ket{0^w}$, and the subleading generalized eigenstates all belong to $V_w$, it is reasonable to assume that the leading singular states reside in this subspace. Thus, we only consider the LLg in this subspace in this section.
The restriction of $T_w$ to $V_w$ can be calculated exactly. Because
\begin{align}
	T_{w=1}^\dag\ket{v}&=\frac{q^2\ket{v}}{q^2+1}\nonumber\\
	\bra{v}\rbra{0}\overline{\UU_2}\ket{v}&=\frac{q\rbra{v}}{q^2+1},
\end{align}
we have for $n\geq 1$ and $m\leq w-2$
\begin{align}
	\mel{e_{m+n}}{T_w}{e_{m}}&=\qty(\frac{q}{q^2+1})^{n-1}\mel{e_{m+1}}{T_w}{e_{m}}\nonumber\\
	&=\qty(\frac{q}{q^2+1})^{n}\frac{q^2}{q^2+1}.
\end{align}
Hence, $T_w$ in subspace $V_w$ can be expressed as
\begin{align}
	T_w'=\left[\begin{array}{ccccccc} 
	z_2 &   &   &   &  & \\ 
	z_2(z_2/q) & z_2  &   &   &  & \\ 
	z_2(z_2/q)^2 &  z_2(z_2/q) & z_2  &   &  &\\ 
	\vdots &  \ddots &  \ddots & \ddots &  &\\
	  z_2(z_2/q)^{w-1} &  \cdots &  z_2(z_2/q)^2 & z_2(z_2/q)  &  z_2  &  \\
	   y_w  & \cdots  & y_3 & y_2 & y_1 & 1 \end{array}\right],
\end{align}
where $z_2=q^2/(q^2+1)$, and $y_i$ are some coefficients to be determined. The matrix can be simplified to
\begin{align}
	T_w'=\left[\begin{array}{ccccccc} 
	R &  0  \\ 
	y &  1\end{array}\right],
\end{align}
where $y=(y_w,y_{w-1},\cdots,y_1)$, $R=[(1+q^{-2})-M/q]^{-1}$ and 
\begin{align}
	M=\left[\begin{array}{ccccccc} 
	0 &   &   &   &   \\ 
	1 & 0  &   &   &   \\ 
	0 &  1 & 0  &   &  \\ 
	\vdots &  \ddots &  \ddots & \ddots & \\
	 0 &  \cdots &  0 & 1  &  0 \end{array}\right].
\end{align}
As $\bra{1^w}\propto[\chi,1]:=[\chi(w),\chi(w-1),\cdots,\chi(1),1]$ is a left eigenstate of $T_w$, we then have
\begin{align}
	[\chi,1]=[\chi,1]T_w'=[\chi R+y,1],
\end{align}
meaning that $y=\chi-\chi R$. Therefore, we obtain
\begin{align}
	F_w'=\left[\begin{array}{ccccccc} 
	R &   0 \\ 
	-\chi R &  0\end{array}\right],
\end{align}
and
\begin{align}
	(F_w')^\tau=\left[\begin{array}{ccccccc} 
	R^\tau &   0 \\ 
	-\chi R^\tau &  0\end{array}\right].
\end{align}
The leading singular value $\lambda_{w,\tau}$ is the square root of leading eigenvalue of 
\begin{align}
	(F_w'^\dag)^\tau(F_w')^\tau=(R^\dag)^\tau R^\tau+(R^\dag)^\tau \chi\chi^\dag R^\tau. 
\end{align}
As $\norm{\chi}=\sqrt{q^{2w}-1}$, the leading eigenvalue of $(F_w'^\dag)^\tau(F_w')^\tau$ is dominated by the second term, and correspondingly, $\lambda_{w,\tau}\approx\norm{\chi R^\tau}$. Further, $R^\tau$ can be calculated from Taylor expansion,
\begin{align}
	R^\tau=\qty(\frac{q^2}{q^2+1})^\tau\sum_{k=0}^{w-1}{{\tau+k-1}\choose{k}}\qty(\frac{q}{q^2+1})^k M^k,
\end{align}
where we use the identity $M^w=0$. The $j$th component of $\chi R^\tau$ is given by
\begin{align}
	[\chi R^\tau]_j&=\qty(\frac{q^2}{q^2+1})^\tau\sum_{k=0}^{j}{{\tau+k-1}\choose{k}}\qty(\frac{q}{q^2+1})^k\nonumber\\
	&\ \ \times \chi(w-k-j)\nonumber\\
	&=\chi(w-j)\sum_{k=0}^{j}{{\tau+k-1}\choose{k}}(1-z_2)^kz_2^{\tau}.
\end{align}
Note that $\sum_{k=0}^{\infty}{{\tau+k-1}\choose{k}}(1-z_2)^k=z_2^{-\tau}$, so $[\chi R^\tau]_j\leq [\chi]_j$.
For large $\tau$ and $j/\tau\approx q^2$, using Stirling's formula, we have
\begin{align}
	[\chi R^\tau]_j&\approx \chi(w-j) \sum_{w'=0}^{j}\sqrt{\frac{z_2}{2\pi(1-z_2)(\tau+w'-1)}}\nonumber\\
	&\ \ \times\exp \qty{\frac{[(1-z_2)w'-z_2(\tau-1)]^2}{2z_2(1-z_2)(w'+\tau-1)}}\nonumber\\
	&\approx  \chi(w-j)f_\tau(j/\tau),
\end{align}
where $f_\tau(x)$ is defined by
\begin{align}
	f_\tau(x)&:=\int_{0}^{x}\dd s\sqrt{\frac{z_2\tau}{2\pi(1-z_2)(1+s)}}\nonumber\\
	&\ \ \times\exp \qty{\frac{[(1-z_2)s-z_2]^2}{2z_2(1-z_2)(1+s)}\tau}\nonumber\\
	&=\int_{0}^{x\tau}\dd w'\sqrt{\frac{z_2}{2\pi(1-z_2)(w'+\tau)}}\nonumber\\
	&\ \ \times\exp \qty{\frac{[(1-z_2)w'-z_2\tau]^2}{2z_2(1-z_2)(\tau+w')}}.
\end{align}
 As $\norm{\chi R^\tau}$ is dominated by components with $j/w\approx 1$, we obtain
 \begin{align}
 	\norm{\chi R^\tau}\approx \norm{\chi}f_\tau(w/\tau)\approx q^w f_\tau(w/\tau).
\end{align}

\section{Additional numerics}
In this section, we provide additional numerics for the behavior of the OTOC around the butterfly cone, the performance of the LSVA and the variation method, and the $w$-dependence of the large $\tau$ behavior in various models.

\subsection{OTOC, LSVA, and the variation method in the spatial-temporal invariant models}
\begin{figure}[h]
\centering
\includegraphics[width=\columnwidth]{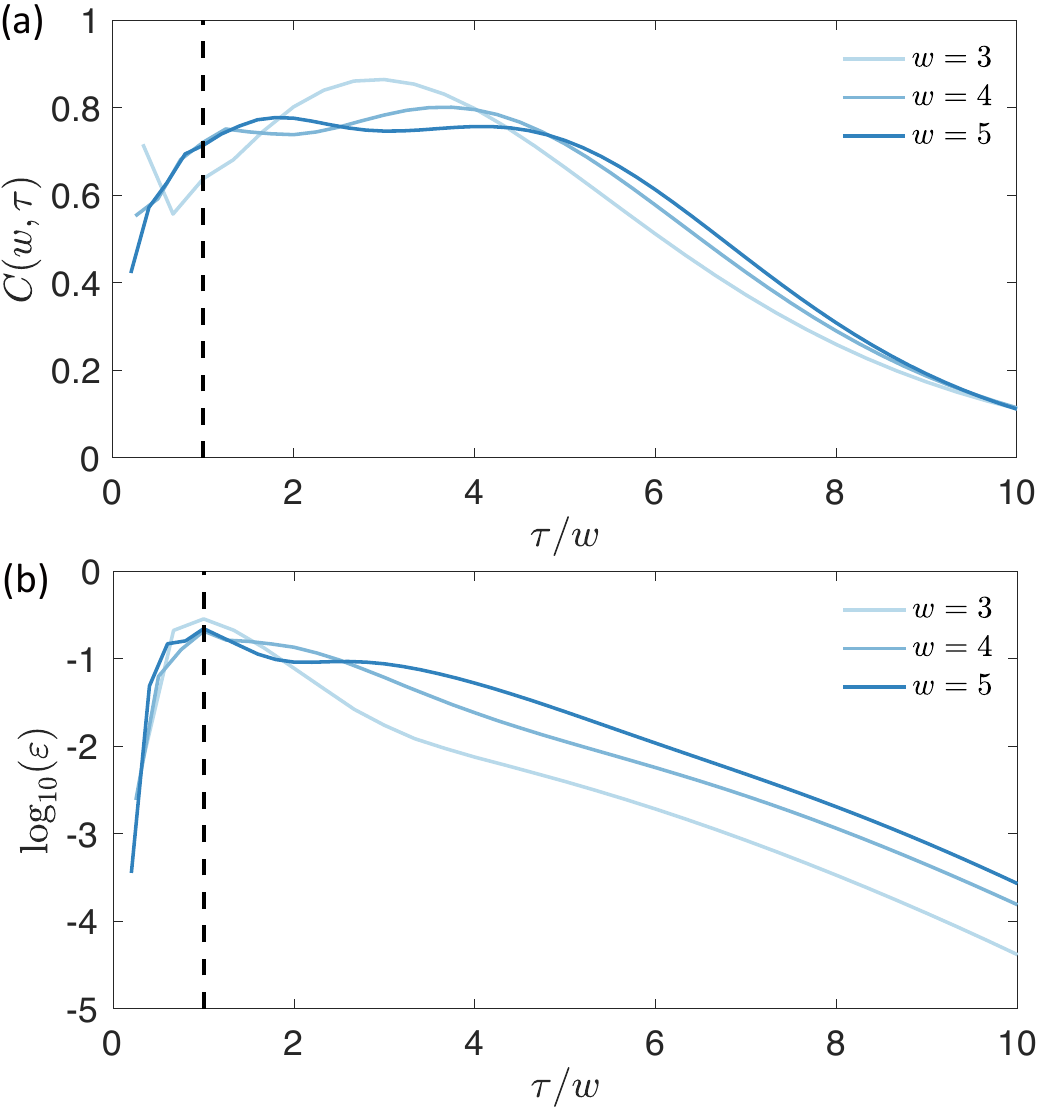}
\caption{(a) OTOC around the butterfly cone and (b) error $\varepsilon=\abs{C-C_{\text{LSVA}}}$ in the spatial-temporal invariant XYZc model. The black dashed lines denote $\tau/w=1$, and we use the same parameters as those in the main text.
\label{Fig:XYZ}}
\end{figure}

\begin{figure}[t]
\centering
\includegraphics[width=\columnwidth]{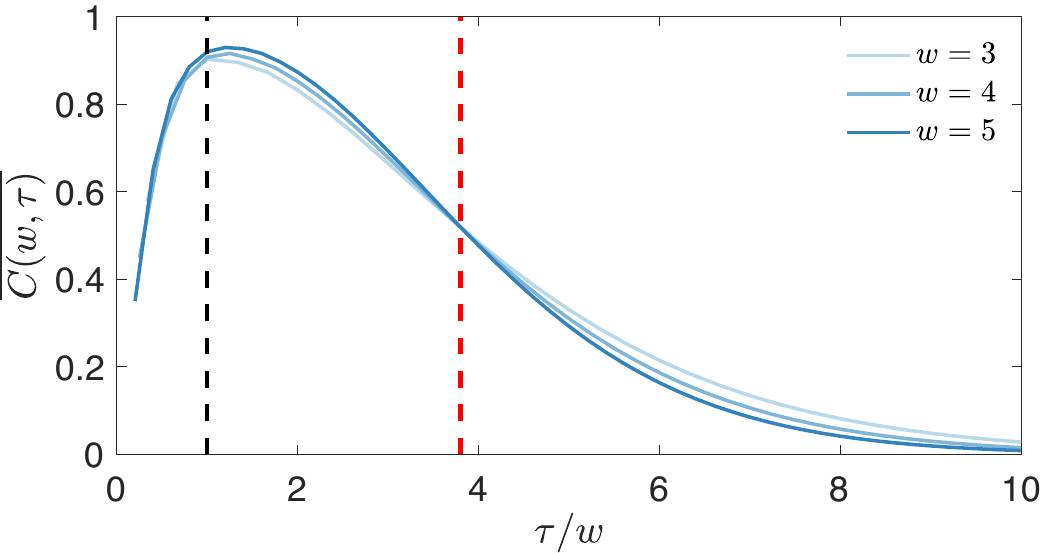}
\caption{(a) Ensemble-averaged OTOC around the butterfly cone in the spatial-temporal invariant 3PM. The black dashed lines denote $\tau/w=1$, and the red ones denote the crossing point of the OTOC at $\tau/w=3.8$. Here, the results are averaged over 25 realizations.
\label{Fig:3PM_EnsembleAve}}
\end{figure}

\begin{figure}[h]
\centering
\includegraphics[width=\columnwidth]{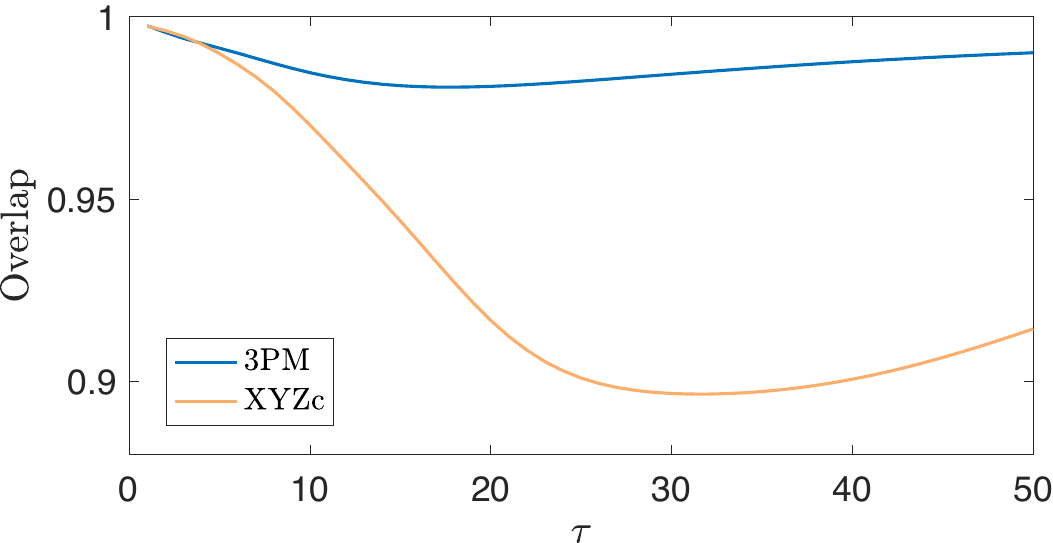}
\caption{Overlap between the exact leading singular state of the left-moving \LLG\ and the variational ansatz, defined by $\abs{\braket{\lambda_{w,\tau}^{\text{L,VP}}}{\lambda_{w,\tau}^\text{L}}\braket{\lambda_{w,\tau}^{\text{R}}}{\lambda_{w,\tau}^\text{R,VP}}}$. Here, we calculate the spatial-temporal invariant case with $w=5$, and use the same realization as that in the main text.
\label{Fig:Overlap}}
\end{figure}

In this subsection, we provide: 
(i) The OTOC and the error of LSVA in the XYZc model in Fig.~\ref{Fig:XYZ},
(ii) the sharpening of the OTOC around the butterfly cone in the 3PM in Fig.~\ref{Fig:3PM_EnsembleAve},
(iii) overlap between the leading singular states and the variational states in the 3PM in Fig.~\ref{Fig:Overlap}.

\subsection{OTOC, LSVA, and the variation method in the spatial-temporal random HRM}
\begin{figure}[H]
\center
\includegraphics[width=0.7\columnwidth]{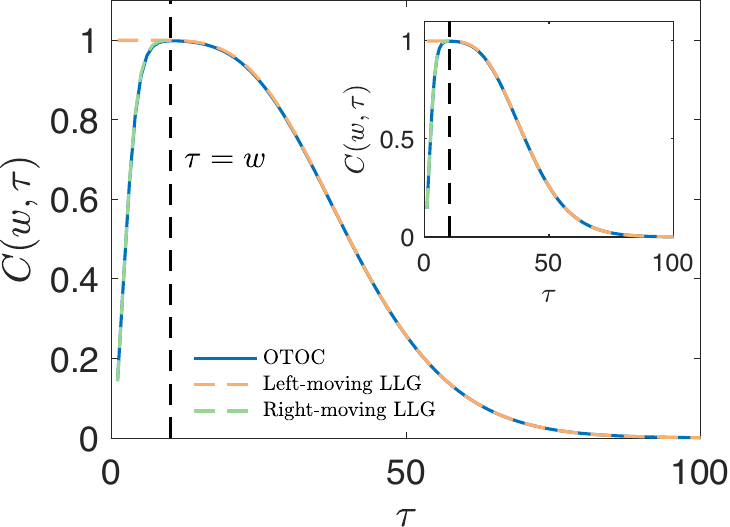}
\caption{\label{Fig:HRM}  OTOC of the spatial-temporal random HRM with $q=2$ and $w=10$. The dashed lines in the main panel and the inset represent the LSVA and the variational method, respectively.}
\end{figure}

\begin{figure}[H]
\centering
\includegraphics[width=\columnwidth]{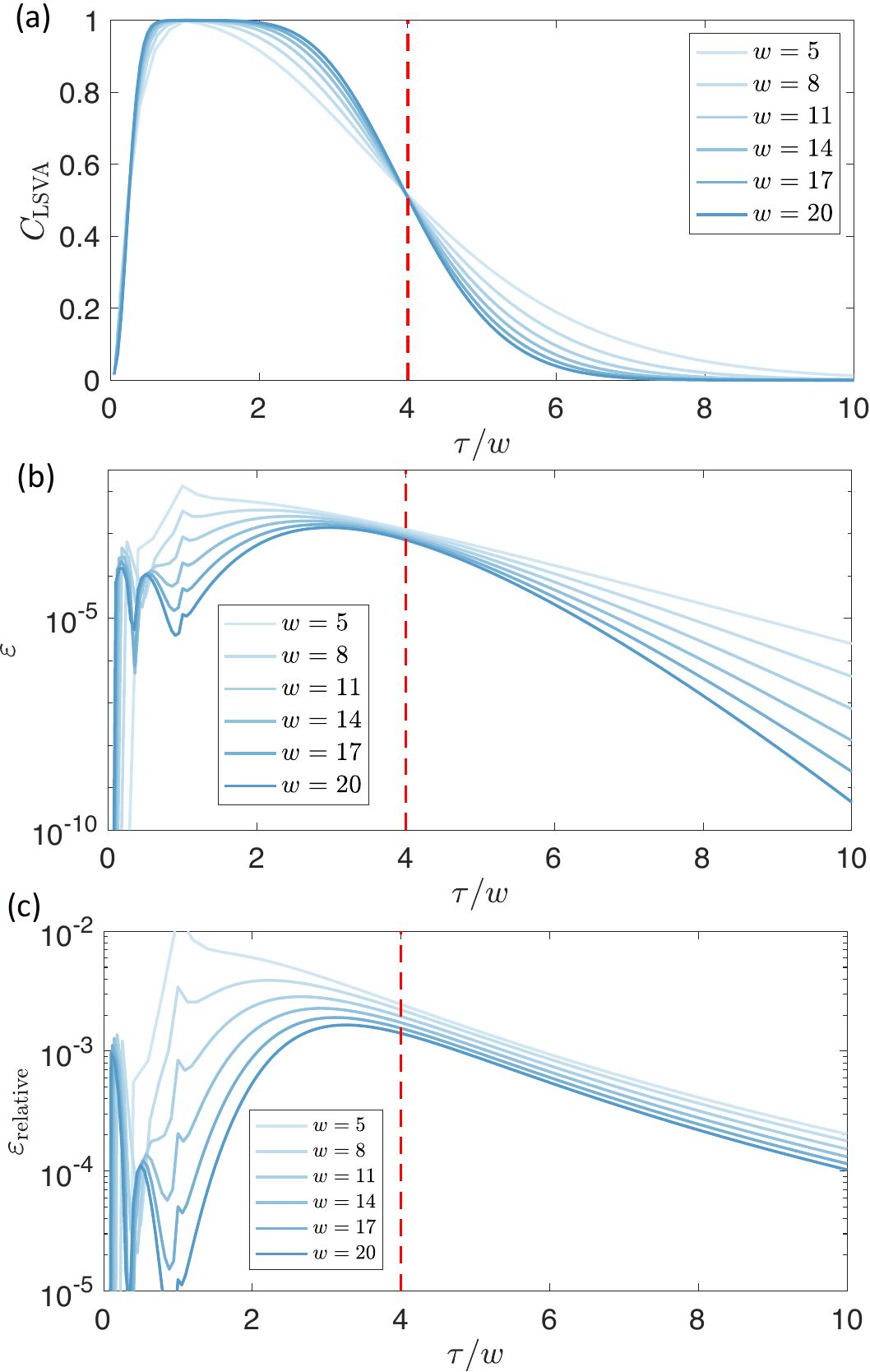}
\caption{(a) LSVA around the butterfly cone, (b) error $\varepsilon=\abs{C-C_{\text{LSVA}}}$, and (c) relative error $\varepsilon_{\text{relative}}=\abs{C-C_{\text{LSVA}}}/C$ in the spatial-temporal random HRM with $q=2$. The red one denotes the position of the butterfly cone $\tau/w=q^2$.
\label{Fig:CaseA_HRM}}
\end{figure}

In this subsection, we provide: 
(i) Comparison between OTOC, LSVA and variational method in this model in Fig.~\ref{Fig:HRM},
(ii) the sharpening of the OTOC around the butterfly cone and the error of LSVA in this model in Fig.~\ref{Fig:CaseA_HRM}.

\subsection{$w$-dependence of the large $\tau$ behavior }
\begin{figure}[H]
\center
\includegraphics[width=\columnwidth]{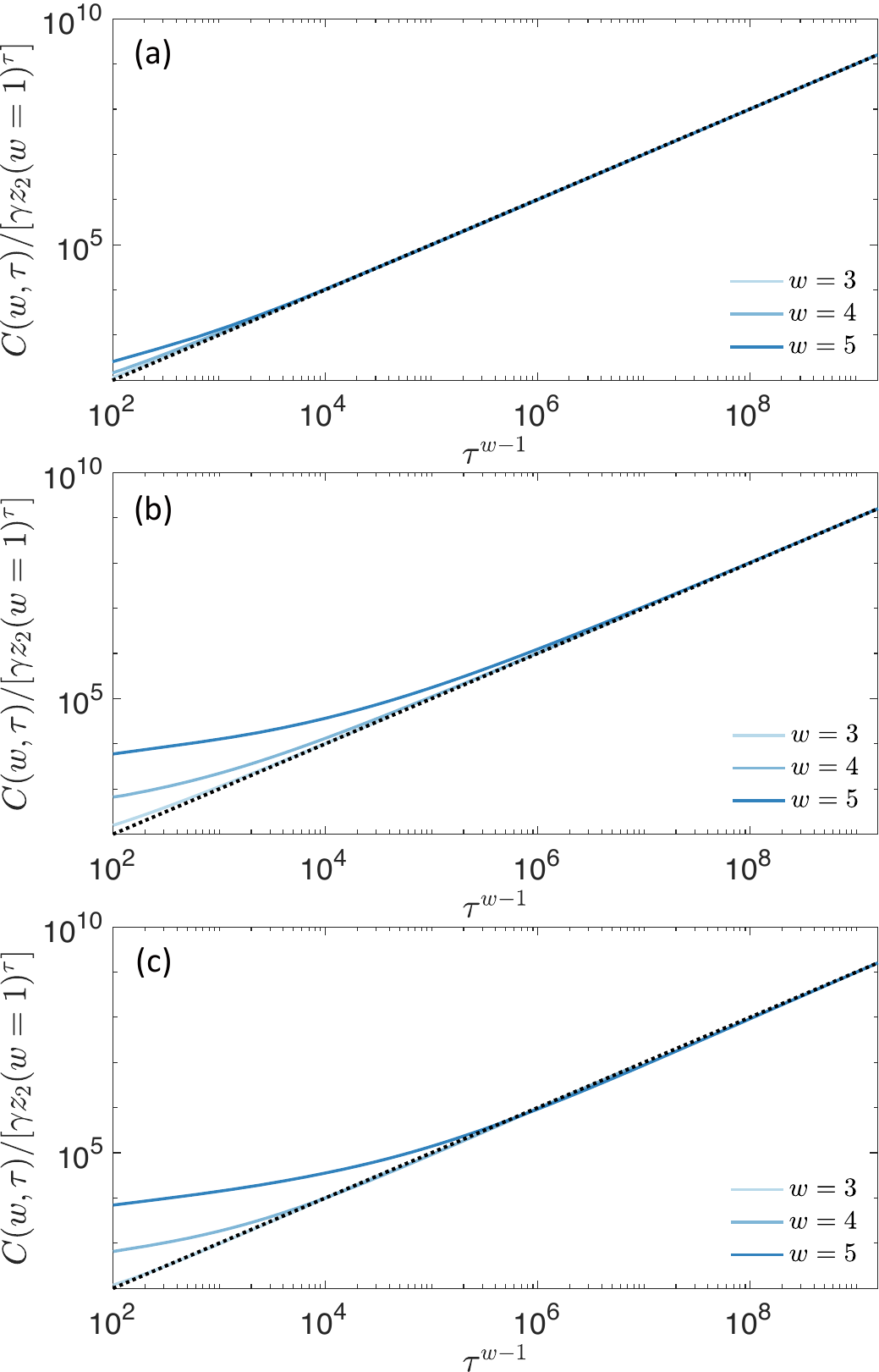}
\caption{\label{Fig:Largetau}Large $\tau$ behavior of the OTOC for (a) spatial-temporal random RPM, (b) spatial-temporal invariant 3PM, and (c) spatial-temporal invariant Z2-COE model. For the spatial-temporal invariant models, we use the same realization as that in the main text.}
\end{figure}

In this subsection, we provide the large $\tau$ behavior for different $w$ in various models in Fig.~\ref{Fig:Largetau}. 

\bibliographystyle{apsrev4-2}
\bibliography{biblio.bib}